\documentclass[superscriptaddress,showkeys]{revtex4-2}
\usepackage{graphicx}
\usepackage{multirow}
\usepackage{lipsum}
\usepackage{placeins}
\usepackage{dcolumn}
\usepackage{bm}
\usepackage[colorlinks=true, citecolor=red, urlcolor = violet, linkcolor= purple, bookmarks=true]{hyperref}
\usepackage{float}
\usepackage{textcomp}
\usepackage[utf8]{inputenc}
\usepackage{siunitx}
\usepackage{amsmath,amsfonts,amssymb}
\usepackage{natbib}
 \usepackage{orcidlink} 
\usepackage{breakurl}
\usepackage{xcolor,verbatim,multirow}
\usepackage[normalem]{ulem}
\usepackage{subcaption} 
\usepackage{wrapfig}
\usepackage{mathrsfs}
\usepackage{pgfplots}
\usepackage{mathtools}
\usepackage{adjustbox}
\usepackage{caption}
\usepackage{booktabs}
\begin{document} 

\title{Testing loop quantum gravity through EHT observations of M87* and Sgr A* using rotating holonomy-corrected black holes}

\author{Heena Ali \orcidlink{0009-0008-2242-7650}}\email{Heenasalroo101@gmail.com}
\affiliation{Centre for Theoretical Physics, Jamia Millia Islamia, New Delhi 110 025, India}
\author{Sushant G. Ghosh\orcidlink{0000-0002-0835-3690}}
\email{sghosh2@jmi.ac.in, sgghosh@gmail.com}
\affiliation{Centre for Theoretical Physics, Jamia Millia Islamia, New Delhi 110 025, India}
\affiliation{Center for Theoretical Physics, Khazar University, 41 Mehseti Street, Baku, AZ1096, Azerbaijan.}
\affiliation{Astrophysics and Cosmology
	Research Unit, School of Mathematics, Statistics and Computer Science, University of
	KwaZulu-Natal, Private Bag 54001, Durban 4000, South Africa}
\begin{abstract}
The Event Horizon Telescope (EHT) has provided a new tool for testing the strong-field regime of gravity by imaging the shadows of M87* and Sgr A*. These observations provide the first real opportunity to test whether quantum gravity, specifically loop quantum gravity, leaves observable imprints on spacetime. We use the EHT observations of M87* and Sgr A* to examine the observational signs of rotating holonomy--corrected black holes (RHCBHs). We discover that, in comparison to the typical Kerr black hole, the quantum correction parameter $b$ increases the size of the black hole shadow. As the deviation parameter $b$ increases in RHCBH, the prograde photon orbits shift outward, indicating a weaker effective gravitational field near the central region. We find that photon rings continue to exist in the parameter range $b_E \leq b \leq b_p$, due to the presence of unstable circular photon orbits. We apply the Kumar--Ghosh method based on the shadow observables: the shadow area $A$ and the oblateness $D$ that together allow a unique determination of the spin parameter $a$ and the quantum correction parameter $b$. 
At $\theta_o=17$\textdegree~, the angular diameter bound of M87$^{*}$ yields $b \leq 0.1319\,M$ at $a = 0\,$ and $b \leq 0.421\,M$ at $a = 0.784\,M$, while at $\theta_o=50$\textdegree~, the angular diameter bound of Sgr A$^{*}$ yields $b \leq 0.5764\,M$ at $a = 0\,$ and $b \leq 0.7482\,M$ at $a = 0.6253\,M$ the Sgr~A$^{*}$. Our results show that nonzero values of the holonomy correction parameter are consistent with current EHT data, indicating that RHCBHs provide viable alternatives to the classical Kerr geometry in the strong-gravity regime and are strong astrophysical black hole candidates. 
\end{abstract}


\maketitle

\section{Introduction}

Black holes are one of the most important grounds for probing General Relativity (GR) \cite{1915SPAW.......844E} in the strong--field regime. The vacuum solutions of the Einstein field equations -- the Schwarzschild metric for a static, spherically symmetric mass distribution and the Kerr metric for a rotating one, both contain curvature singularities, as established by the singularity theorems \cite{Penrose:1964wq,Hawking:1970zqf}. This signals the breakdown of classical GR at Planck-scale energies and the need to modify the classical description in regimes of higher curvature, thereby motivating a more fundamental description of the quantum theory of gravity that supports the quantization of spacetime near the gravitational singularities. Loop quantum gravity (LQG) emerges as a promising theory in such a case, addressing this problem through a non-perturbative, background-independent approach of quantizing the gravitational field. The field of loop quantum gravity (LQG) is a non-perturbative, background-independent approach to quantize the gravitational field, replacing the smooth continuum of general relativity with a discrete structure at the Planck scale. Ashtekar, Rovelli, and Smolin \cite{Ashtekar:1992tm} established the foundational framework of loop quantum gravity, showing how classical spacetime geometry can emerge from an underlying quantum description and connecting LQG to classical geometry at macroscopic scales. Later, Rovelli \cite{Rovelli:2004tv} gave a detailed exposition of the canonical formulation of LQG, including the quantisation of the gravitational field and the regularisation of the Hamiltonian constraint. The spin-foam approach provides the covariant formulation of LQG, was comprehensively reviewed by Perez \cite{Perez:2012wv}, establishing an important foundation for the development of quantum gravity phenomenology. The more recent textbook by Rovelli and Vidotto \cite{Rovelli:2014ssa} presents the covariant formulation of LQG, including the spinfoam theory that underlies many effective models of quantum black holes. It replaces the smooth continuum of GR by a discrete structure at the Planck scale \cite{Rovelli:2004tv, Ashtekar:2004eh} with the objective of regularizing gravity \cite{Soares:2023uup}. A distinguishing feature of LQG is that it replaces the curvature components in the classical Hamiltonian constraint by finite corrections built from $SU(2)$ holonomies along closed loops \cite{Bojowald:2002gz, Ashtekar:2006uz}. Consequently, these corrections introduce the minimum eigenvalue of the area operator, $\Delta=4\sqrt{3}\pi\gamma\ell_{\mathrm{Pl}}^{2}$, $\gamma$ being the Barbero--Immirzi parameter regulating the classical curvature divergences that would otherwise be inevitable at $r \to 0$ \cite{Modesto:2005zm, Ashtekar:2005qt}.

Upon applying LQG techniques to spherically symmetric black holes, a class of effective metrics with the classical Schwarzschild singularity replaced by a quantum bounce is yielded \cite{Modesto:2005zm, Gambini:2008wn, Boehmer:2007ket}. The resulting spacetime is termed the loop quantum black hole (LQBH). Throughout the maximally extended manifold, the Kretschmann scalar remains finite, and a second asymptotically flat region emerges on the other side of the bounce \cite{Modesto_2009, Caravelli:2010ff}. This deviation of behaviour from the holonomy-corrected geometries of classical black holes provides concrete motivation to test these geometries observationally, and thus their study has attracted considerable attention \cite{Alonso-Bardaji:2022ear,Alonso-Bardaji:2024tvp,Belfaqih:2024vfk,Soares:2023uup}.

To connect LQG with real astrophysical observations, it is important to study rotating quantum-corrected black holes, since most black holes in the universe are expected to rotate and carry significant angular momentum. A well-known recipe for obtaining rotating black hole metrics from static and spherically symmetric spacetimes is the Newman–Janis algorithm (NJA) \cite{Newman:1965tw}. However, the standard complexification procedure in the NJA often yields rotating metrics that violate energy conditions or exhibit pathological behaviour. To circumvent these issues, Azreg-A\"inou's non-complexification procedure offers a systematic and physically well-motivated framework for generating rotating metrics while preserving desirable physical properties \cite{Azreg-Ainou:2014aqa, Azreg-Ainou:2014pra}. The modified Newman-Janis algorithm (MNJA)  provides a systematic procedure for generating rotating spacetimes from static seed metrics, yielding Kerr-like solutions that carry the spin parameter $a$ alongside the quantum correction parameters \cite{Brahma:2020eos, Islam:2022wck, Liu:2020ola}. 

The shadow cast by a black hole provides a direct observational probe of geodesic motion in the strong-field regime. The shadow boundary is determined by a set of unstable circular null geodesics that form the photon sphere (or photon region, in the rotating case) \cite{Synge:1966, Bardeen1973, Luminet:1979nyg}. Any kind of departure from Kerr geometry, be it size, oblateness, or displacement, encodes an imprint of the underlying gravitational theory and can thus be used to constrain modifications to the Einstein--Hilbert action \cite{Bambi:2019tjh, Vagnozzi:2022moj}.
A number of recent studies have explored black hole shadows as a possible probe of LQG and related quantum-inspired theories. The possibility of testing LQG through astrophysical observations has gained significant momentum following the Event Horizon Telescope (EHT) observations of M87* and Sgr A*. A landmark paper by Brahma, Chen, and Yeom \cite{Brahma:2020eos} argued that nonsingular rotating black holes from LQG leave observable footprints that can be tested using current observations. Their work used the MNJA to construct a rotating LQG black hole from a spherically symmetric seed metric—the exact methodology we employ. The possibility of testing LQG through astrophysical observations has gained significant momentum following the Event Horizon Telescope (EHT) observations of M87* and Sgr A*. A landmark paper by Brahma, Chen, and Yeom \cite{Brahma:2020eos} argued that nonsingular rotating black holes from LQG leave observable footprints that can be tested using current observations. Their work, published in \textit{Physical Review Letters}, used the Modified Newman-Janis Algorithm (MNJA) to construct a rotating LQG black hole from a spherically symmetric seed metric—the exact methodology we employ in our work. This demonstrates that our approach is an established and accepted method for connecting LQG with astrophysical observations. The MNJA methodology developed by Azreg-Aïnou \cite{Azreg-Ainou:2014aqa,Azreg-Ainou:2014pra} to address ambiguities in the complexification procedure of the standard Newman-Janis algorithm, has been extensively employed in the LQG phenomenology literature. Liu et al. \cite{Liu:2020ola} applied MNJA to construct a rotating LQG black hole to discuss its shadows and quasinormal modes, showing that LQG corrections can leave potentially observable signatures. This line of work was subsequently extended to observational tests. Islam {\it et al.} \cite{Islam:2022wck}, for example, confronted rotating LQG black holes with EHT observations and obtained constraints on the LQG parameter from the shadow of the black hole. More recently, Ali {\it et al.} \cite{Ali:2024ssf} considered rotating quantum-corrected black holes in connection with EHT observations of both M87* and Sgr A*. Similarly, Kumar {\it et al.} \cite{Kumar:2022vfg} examined multi-horizon LQG motivated rotating black holes LQG, further illustrating the use of the MNJA in constructing and studying rotating LQG-inspired geometries. All these works use the MNJA to generate rotating geometries from LQG-inspired static seeds and classify them as ``LQG-motivated'' or ``LQG-inspired'' black holes, establishing this as the standard phenomenological approach in the field. One cluster of studies focused specifically on rotating LQG black holes. Liu and his collaborators \cite{Liu:2020ola} were among the first to compute shadows and quasinormal modes for such objects. Around the same time, Brahma, Chen, and Yeom \cite{Brahma:2020eos} argued that nonsingular rotating black holes from LQG leave observable footprints that we can actually test today. Following up on this, Kumar Walia \cite{KumarWalia:2022ddq} made observational predictions for polymerized black holes and used EHT data from both Sgr A* and M87* to put preliminary constraints on the model. Afrin {\it et al.} \cite{Afrin:2022ztr} then performed a more rigorous LQG test using the Sgr A* shadow, showing that certain quantum parameters can be strongly constrained. \cite{Islam:2022wck} extended this analysis to rotating black holes, while Kumar {\it et al}\cite{Kumar:2023jgh} looked at strong gravitational lensing in the same context. Jiang {\it et al.} \cite{Jiang:2023img}  ran semi-analytical simulations tailored specifically for Sgr A* and M87*, giving us detailed predictions of what the shadows should look like. Sekhmani {\it et al.}   \cite{Sekhmani:2025bsi}   took the LQG framework a step further by adding electric charge and studied the implications of EHT on nonsingular rotating charged black holes.

 Any modification to the photon capture cross-section relative to Kerr is quantitatively constrained by the EHT measurements for the angular diameter of the shadow and parameterised through the Schwarzschild shadow deviation \cite{Vagnozzi:2022moj}. When the interplay between the holonomy correction parameter and angular momentum is such that the formation of an event horizon is prevented, the geometry of spacetime transitions into a regular compact object. Whether such a horizonless configuration can be distinguished observationally from a Kerr black hole of comparable mass and spin using EHT data is still a well-motivated open question. These EHT observations provide an exceptional opportunity to test modified theories of gravity in the strong-field regime, where deviations from classical GR are expected.  
The close agreement between the observed black hole shadow and the predictions of the Kerr black hole model places strong constraints on alternative theories of gravity. At the same time, these observations also provide a unique opportunity to study possible quantum gravity effects near the event horizon \cite{EventHorizonTelescope:2020qrl, EventHorizonTelescope:2021dqv, Vagnozzi:2022moj}. The EHT observations have already been used to test several modified gravity models \cite{Cunha:2019dwb, Stuchlik:2019uvf, Jusufi:2022loj}. Interpreting EHT observations requires a reliable theoretical framework to understand black hole shadows and photon rings. Psaltis {\it et al.} \cite{Psaltis:2014mca} established a general relativistic null hypothesis test for EHT observations of the black hole shadow in Sgr A*, providing the foundation for constraining modifications to general relativity using shadow measurements. In contrast, the first EHT observations of M87* demonstrated the potential of horizon-scale imaging to test strong-field gravity \cite{EventHorizonTelescope:2019dse, EventHorizonTelescope:2019ggy}. Psaltis {\it et al.} \cite{Psaltis:2020lvx} further showed that the M87* shadow can constrain deviations from the Kerr geometry. Gralla, Holz, and Wald \cite{Gralla:2019xty} studied the relation between black hole shadows, photon rings, and gravitational lensing. They showed that the shadow boundary is determined by the unstable photon orbits in the photon region and is not directly tied to the presence of an event horizon. This is in line with our result that a closed shadow can also be obtained in spacetimes without an event horizon. Similar possibilities have been explored for black hole mimickers and modified-gravity models using EHT observations \cite{Vagnozzi:2022moj}. In models motivated by quantum gravity, the size of the black hole shadow can give important information about the quantum correction parameters and may help us understand possible deviations from classical gravity \cite{Eichhorn:2019dhg, Jusufi:2020zln, Yang:2022btw}.

Motivated by the above, we study the signatures of rotating holonomy-corrected black holes (RHCBHs) inspired by loop quantum gravity (LQG). We investigate how the quantum correction parameter $b$ changes the photon region and the black hole shadow compared to the Kerr black hole. Using the Event Horizon Telescope (EHT) observations of M87* and Sgr A*, we constrain the RHCBH parameters from the observed shadow size and Schwarzschild deviation parameter. We also use the shadow area and oblateness to estimate the black hole parameters. In addition, we find that the RHCBH spacetime can still produce closed photon rings even when no event horizon is present.

The paper is organized as follows. In Section~\ref{sect2}, we revisit the static holonomy-corrected Schwarzschild black hole that serves as our seed metric, apply the MNJA to derive the RHCBH spacetime, and analyze its horizon structure. ~\ref{sect3} is dedicated to the study of photon motion; we derive the null geodesic equations via the Hamilton-Jacobi formalism, establish the conditions for the existence of the photon region, and investigate the unique closed shadow configurations present in the horizonless spacetime of the parameter space. In ~\ref{sect4}, we introduce the Kumar-Ghosh method based on the shadow area ($A$) and oblateness ($D$) observables and outline a systematic framework for black hole parameter estimation. Section~~\ref{sect5} applies rigorous observational constraints on the RHCBH parameter space by matching theoretical predictions against the shadow angular diameters and Schwarzschild deviation limits reported by the Event Horizon Telescope for $\text{M87}^*$ and $\text{Sgr A}^*$. Finally, we compare our results with other well-known rotating and regular black hole metrics in Section~~\ref{sect6}, and summarize our core conclusions and future outlook in Section~~\ref{sect7}.

\section{Holonomy corrected black holes}\label{sect2}
Before proceeding to the rotating case, we briefly review the static holonomy-corrected Schwarzschild black hole and its applications.  
Several researchers have looked at the non-rotating, spherical version of these holonomy-corrected black holes. Alonso-Bardaji and his collaborators \cite{Alonso-Bardaji:2022ear, Alonso-Bardaji:2023niu, Alonso-Bardaji:2024tvp} have done a series of studies on this topic. They first built a nonsingular spherical black hole model with holonomy corrections, then added electric charge to see how it behaves in cosmological backgrounds, and later studied how such black holes actually form.
 Moreira {\it et al} \cite{Moreira:2023cxy} calculated quasinormal modes for a holonomy-corrected Schwarzschild black hole. Bolokhov \cite{Bolokhov:2023bwm} went a step further and looked at long-lived modes and overtones. Gingrich \cite{Gingrich:2024tuf} did a similar analysis for the nonsingular spherical model. Yang and his team \cite{Yang:2024ofe} took a wider view and studied both quasinormal frequencies and Hawking radiation together. Junior {\it et al}\cite{Junior:2023xgl} studied lensing for the holonomy-corrected Schwarzschild black hole. Soares and collaborators \cite{Soares:2023uup, Soares:2024rhp} first looked at basic lensing and later included topological charge in their calculations. 
Belfaqih {\it et al.} \cite{Belfaqih:2024vfk} worked on making holonomy corrections covariant, which is important for keeping the theory consistent with GR. Huang \cite{Huang:2025vpi} studied how holonomy corrections change the orbits of objects moving around the black hole, paying special attention to precessing and periodic orbits. 
\subsection{Holonomy corrected Schwarzschild black holes } \label{HCBH}
Now, let us briefly review the holonomy-corrected Schwarzschild black hole, which will serve as the seed metric for our rotating metric. The line element describing this static, spherically symmetric spacetime reads \cite{Alonso-Bardaji:2021yls, Alonso-Bardaji:2022ear, Soares:2023uup}
\begin{eqnarray}\label{sphericalmetric}
ds^2&=&\bigg(1-\frac{2M}{r}\bigg)dt^2-\bigg(\frac{r}{r-b}\bigg)\bigg(1-\frac{2M}{r}\bigg) ^{-1}dr^2\nonumber\\
&& - r^2\left(d\theta^2+\sin^2\theta d\phi^2\right) \ .
\end{eqnarray}
Here, the LQG parameter $b$, having dimensions of length, incorporates the effects of  quantum gravity through the dimensionless holonomy parameter $\lambda$, \cite{Huang:2025vpi}
\begin{eqnarray}\label{b}
b=r_s \frac{\lambda^2}{\lambda^2+1},
\end{eqnarray}
where $r_s\equiv {2GM}/{c^2}$ is the Schwarzschild radius. 
The metric~(\ref{sphericalmetric}) is characterised by a wormhole-like structure, where the trapped regular BH interior is separated from the anti-trapped other region by a minimal spacelike hypersurface defined by $b$ \cite{Soares:2023uup}. However, we shall take into account light rays that do not pass through the event horizon, denoted by $r = r_h = 2M$, i.e., we will consider regions where $r > r_h$.
As evident from Eq.~\ref{b}, the LQG parameter $b$ is restricted to the range $b\in [0,r_s]$. The metric (\ref{sphericalmetric}) reduces to the Schwarzschild black hole upon neglecting the contribution of $b$. When $0 < b < r_s$, the black hole has both an inner (Cauchy) horizon and an outer (event) horizon. When $b=0$ or $b=r_s$, only one horizon exists \cite{Huang:2025vpi}. We cannot differentiate the holonomy-corrected Schwarzschild black hole from the Schwarzschild black hole on the basis of the event horizon, as the event horizons of both always coincide at $r_s$.
\begin{figure*}[hbt!]
\centering
\includegraphics[scale=0.9]{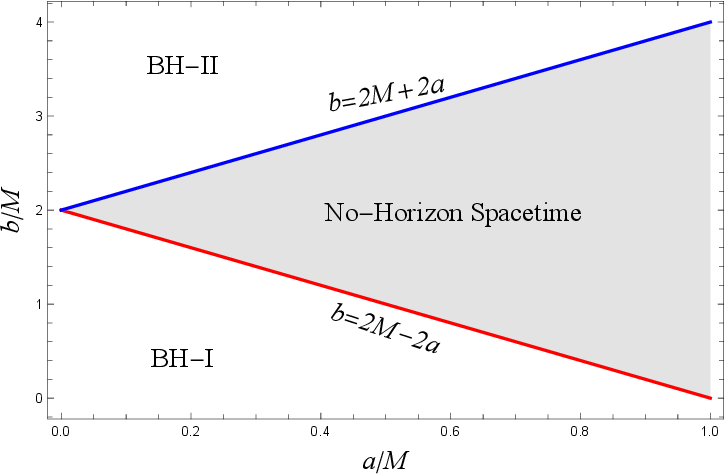}
\caption{\label{HCPS}The parameter space $(a, b)$ for the RHCBH metric. The blue and red lines, respectively, represent the extremal black holes' boundary with degenerate horizons defined by the equation $b = 2M \pm 2a$. The central grey shaded region corresponds to a no-horizon spacetime where the metric function $\Delta(r)$ possesses no real roots, describing a regular, horizonless compact object. The unshaded domains correspond to the physically admissible black hole phases BH-I ($b \le 2M - 2a$) and BH-II ($b \ge 2M + 2a$).}
\end{figure*}

The holonomy-corrected Schwarzschild seed metric resolves the classical singularity by replacing the region $r < b$ with a minimal spacelike hypersurface at $r = b$. The Kretschmann scalar is finite at the minimal surface, $K(b) = 16M^2/b^6$, confirming the regularity of the spacetime in the observational region $r \geq b$. For $r < b$, the spacetime is not part of the physical manifold, and geodesics are extended through the bounce into a white hole region. Thus, the spacetime is geodesically complete and regular in the sense relevant for the EHT observations, which probe the exterior region $r \geq 2M$.

\subsection{Rotating spacetime metric}

Given the above holonomy-corrected Schwarzschild metric, we seek its rotating counterpart using the MNJA obtained by Azreg-A\"inou’s non-complexification procedure \cite{Azreg-Ainou:2014aqa,Azreg-Ainou:2014pra}, which in Boyer-Lindquist coordinates  ($t, r, \theta, \phi$) reads as
\begin{eqnarray}\label{rotmetric}
ds^2&=&\frac{\Psi}{\rho^2}\Big[\frac{\Delta}{\rho^2}(dt-a\sin^2\theta d \phi)^2-\frac{\rho^2}{\Delta}\,d r^2-\rho^2 d \theta^2\nonumber\\
&& -\frac{\sin^2\theta}{\rho^2}\,[ad t-(K+a^2)d\phi]^2\Big],
\end{eqnarray}
where 
\begin{align}\label{rho}
&\rho^2= K+a^2 \cos^2\theta,\;\;K=\sqrt{1-\frac{b}{r}} r^2,\nonumber\\
&\Delta(r)= r^2-(2M+b)r+a^2+2Mb.
\end{align}
$a=L/M$ is the rotation parameter, and $M$ and $L$ are the mass and angular momentum components, respectively. The LQG parameter $b$ measures the potential deviation of the metric~(\ref{rotmetric}) from the standard Kerr black hole metric. It can be easily shown that, in metric~(\ref{rotmetric}), by putting $b=0$, we retrieve the Kerr black hole (\cite{Kerr:1963ud}) and at $a=0$, it reverts to the metric~(\ref{sphericalmetric}). We emphasize that the rotating metric obtained via the MNJA should be interpreted as an effective, phenomenologically motivated geometry that encodes the key holonomy corrections expected from LQG. It is not derived from a full quantization of the rotating LQG Hamiltonian constraint since such solutions remain unavailable. This approach is standard in the field and follows the methodology established in \cite{Brahma:2020eos}, where the same procedure was used to construct a rotating LQG black hole for observational testing. We therefore refer to the spacetime~(\ref{rotmetric}) as a rotating holonomy-corrected black hole to emphasize its effective nature while clearly indicating its LQG motivation and maintain definiteness.

The function $\Psi(r,\theta, a)$ in metric~(\ref{rotmetric}) remains unknown at this stage, but there are certain criteria that may determine this function. For example, according to Ref. \cite{Azreg-Ainou:2014nra}, the following nonlinear differential equation is satisfied by the function $\Psi$.
\begin{equation}
\left(K+a^2 y^2\right)^2\bigl(3\partial_r\Psi\partial_{yy}\Psi-2\Psi\partial_{ryy}\Psi\bigr) = 3a^2\partial_r K\Psi^2,
    \label{eq:Psi_cond_1}
\end{equation}
where $y\equiv\cos\theta$. For an imperfect fluid rotating about the $z$-axis, the aforementioned condition guarantees that the $r\theta$-component of the Einstein tensor will vanish.

The event horizon is defined by the surface $g^{rr} =\Delta= 0$, which yields
\begin{equation}
r^2-(2M+b)r+a^2+2Mb=0.
\label{horizon}
\end{equation}

Eq.~(\ref{horizon}), depending on the values of $a$ and $b$, admits a maximum of two unique real positive roots $(r\pm)$, equal or no real roots, where $r_+$ indicates the outer (event) horizon and $r_-$ indicates the inner (Cauchy) horizon.
\begin{equation}
r\pm=\bigg(M+\frac{b}{2}\bigg)\pm \sqrt{\bigg(M+\frac{b}{2}\bigg)^2-\bigg(a^2+2Mb\bigg)}.
\label{horizonroots}
\end{equation}

The mass $M$, the angular momentum $a$, and the parameter $b$ originating from LQG determine whether the horizon and its radii exist for the RHCBH. Here in this case, for horizons to exist, Eq.~(\ref{horizonroots}) requires 
\begin{equation}
\label{horizon2}
\bigg(M+\frac{b}{2}\bigg)\geq\sqrt{a^2+2Mb}.
\end{equation}

 The non-extremal RHCBH is represented by the inner and outer horizons, respectively, when Eq. (\ref{horizon}) has two distinct real roots. For $a$ and $b$, specific values $a_E$ and $b_E$ exist, where the double real root of Eq. (\ref{horizon}) corresponds to an extremal black hole, given by
\begin{align}\label{extremal}
&\bigg(M+\frac{b_E}{2}\bigg)^2-\bigg(a_E^2+2Mb_E\bigg)=0. \nonumber\\
& a_E=\sqrt{\bigg(M+\frac{b_E}{2}\bigg)^2-2Mb_E}.
\end{align}
Hence, for the extremal case,
\begin{equation}
r_E=M+\frac{b_E}{2}.
\label{rextremal}
\end{equation}
For $b_E>0$, $a_E\leq1$; for the Kerr case, when $b_E=0$, $a_E=1$. For $b_E<0$, $a_E>1$ e.g., for $b_E=-1, a=1.5$. But this case is ruled out in our analysis. The different values of $b_E/M,  a_E/M$ and $r_E/M$ for an extremal  RHCBH are given in Table~\ref{table1A}.

The extremal RHCBH is characterized by the coincidence of the event horizon and the Cauchy horizon determined by putting the discriminant, $D=0$, of $\Delta(r)=0$, solving which gives the following
\begin{equation}
b=2M\pm2a.
\label{alpha}
\end{equation}

The fact that Eq. (\ref{horizon}) does not accept any real root for certain values of $a$ and $b$ is another intriguing example. Since the rotating spacetime solution in this instance represents a normal spacetime solution without a horizon, as we will demonstrate later, it is referred to as the ` "regular non-black-hole solution" \cite{Azreg-Ainou:2014pra,Huang:2025vpi}.  The parameter space $(a,b)$ is shown in Fig.~(\ref{HCPS}). The two straight lines $b=2M\pm2a$ demarcate the boundary between a black hole and a spacetime with no-horizon. For $b>2M-2a$ ans $b<2M+2a$ (grey region in Fig.~\ref{HCPS}), $D<0$, indicating the absence of horizons and hence no-horizon spacetime (cf.~Fig.~\ref{HCPS}). Similar analysis of extremal black holes has been widely investigated for rotating black holes \cite{Bardeen:1972fi, Carter:1968rr, Ghosh:2014pba, Johannsen:2013szh}.
The physically admissible black hole solution exists for $b\leq2M-2a$ (BH-I) and $b\geq2M+2a$ (BH-II) while the region $b\leq2M+2a$ is ruled out since Eq.~\ref{b} constrains $b\leq2M$ \cite{Huang:2025vpi}. When $b=0$, $\Delta=0$ yields
\begin{eqnarray} r_{\pm}^{k} = M\pm\sqrt{M^2-a^2}, 
\end{eqnarray} 
$r_{\pm}^{k}$ are the horizons associated with the Kerr black hole when $a \leq M$.
\section{Photon motion and black hole shadows  }\label{sect3}
We determine the photon motion in the RHCBH spacetime using Hamilton$-$Jacobi equation given below \cite{Carter:1968rr} .
\begin{eqnarray}
\label{HmaJam}
\frac{\partial S}{\partial \tau} = -\frac{1}{2}g^{\alpha\beta}\frac{\partial S}{\partial x^\alpha}\frac{\partial S}{\partial x^\beta} .
\end{eqnarray}
With the help of Eq.~(\ref{HmaJam}), taking  $\tau$ as the affine parameter along the geodesics and $S$ as the Jacobi action, we can derive the corresponding geodesic equations given by
\begin{eqnarray}
S=-{\cal E} t +{\cal L} \phi +S_r(r)+S_\theta(\theta) \label{action},
\end{eqnarray}
with $S_r(r)$ and $S_{\theta}(\theta)$ being the functions only of the $r$ and $\theta$ coordinates respectively. Since the metric (\ref{rotmetric}) has time-- translation and rotational symmetry, it leads to two conserved quantities;  energy $\mathcal{E}=-p_t$ and axial angular momentum $\mathcal{L}=p_{\phi}$ with $p_{\mu}$ being the photon's four-momentum. 
The separability of the Hamilton--Jacobi equation further implies the existence of an additional conserved quantity, namely the Carter constant, which characterizes the latitudinal motion of photons \cite{Carter:1968rr}. Consequently, the null geodesics in the RHCBH spacetime can be expressed in terms of a set of first-order differential equations. These equations completely determine the photon dynamics and play a crucial role in studying important optical properties of the spacetime, such as black hole shadows, photon rings, gravitational lensing, and the structure of the photon region \cite{Bardeen:1973gs,Teo:2003ltt,Perlick:2021aok}. The resulting null geodesic equations in first-order form are written as \cite{Carter:1968rr,Chandrasekhar:1985kt}
\begin{eqnarray}
\Sigma \frac{dt}{d\tau}&=&\frac{r^2+a^2}{\Delta}\left({\cal E}(r^2+a^2)-a{\cal L}\right)  -a(a{\cal E}\sin^2\theta-{\mathcal {L}})\ ,\label{tuch}\\
\Sigma \frac{dr}{d\tau}&=&\pm\sqrt{\mathcal{V}_r(r)}\ ,\label{r}\\
\Sigma \frac{d\theta}{d\tau}&=&\pm\sqrt{\mathcal{V}_{\theta}(\theta)}\ ,\label{th}\\
\Sigma \frac{d\phi}{d\tau}&=&\frac{a}{\Delta}\left({\cal E}(r^2+a^2)-a{\cal L}\right)-\left(a{\cal E}-\frac{{\cal L}}{\sin^2\theta}\right)\ ,\label{phiuch}
\end{eqnarray}
where 
$\mathcal{V}_r(r)$ and $\mathcal{V}_{\theta}(\theta)$ are the effective potentials for radial and polar motion respectively, and are given by:   
\begin{eqnarray}\label{06}
\mathcal{V}_r(r)&=&\left[(K+a^2){\cal E}-a{\cal L}\right]^2-\Delta[{\cal K}+(a{\cal E}-{\cal L})^2]\label{rpot},\quad \\ 
\mathcal{V}_{\theta}(\theta)&=&{\cal K}-\left[\frac{{\cal L}^2}{\sin^2\theta}-a^2 {\cal E}^2\right]\cos^2\theta.\label{theta0}
\end{eqnarray}
Here, ${\cal K}$ is the separation constant emerging from the separability of the Hamilton--Jacobi equation and is closely related to the hidden symmetries of the spacetime. Following the recipe in \cite{Carter:1968rr}, the Carter constant ${\cal Q}$ can be expressed as 
${\cal Q} = {\cal K} + (a{\cal E}-{\cal L})^2,$
which reflects the existence of a non-trivial second-order Killing tensor associated with the RHCBH geometry. The presence of this additional conserved quantity guarantees the complete integrability of photon motion in the rotating spacetime \cite{Frolov:2017kze}. For convenience, and to simplify the analysis of photon trajectories and black hole shadows, it is useful to introduce the dimensionless impact parameters \cite{Chandrasekhar:1985kt,Bardeen:1973gs} \begin{equation} \xi \equiv \frac{\mathcal{L}}{\mathcal{E}}, \qquad \eta \equiv \frac{\mathcal{K}}{\mathcal{E}^2}, \end{equation} where $\xi$ and $\eta$ characterize, respectively, the apparent azimuthal angular momentum and the latitudinal motion of photons as observed at infinity. These parameters effectively reduce the degrees of freedom of the null geodesics from three to two and play a central role in determining the photon region, unstable circular photon orbits, and the contour of the black hole shadow \cite{Bardeen:1973gs,Teo:2003ltt,Perlick:2021aok,Grenzebach:2014fha}.

The nature of photon trajectories in the RHCBH spacetime is completely determined by the radial effective potential $\mathcal{V}_r(r)$ together with the conserved quantities associated with the null geodesics. Depending on the values of the impact parameters and the spacetime geometry, photons may either escape to spatial infinity, fall into the black hole, or remain trapped in unstable bound orbits around the compact object. Of particular importance are the unstable circular photon orbits, as they form the boundary of the black hole shadow perceived by a distant observer \cite{Bardeen:1973gs,Chandrasekhar:1985kt,Perlick:2021aok}.
The conditions for the existence of unstable spherical photon orbits at $r=r_p$ are obtained from the radial potential through
\begin{equation}
\left.\mathcal{V}_r\right|_{(r=r_p)}=\left.\frac{\partial \mathcal{V}_r}{\partial r}\right|_{(r=r_p)}=0,\,\, \text{and}\,\, \left.\frac{\partial^2 \mathcal{V}_r}{\partial r^2}\right|_{(r=r_p)}> 0,\label{vr} 
\end{equation}
$r_p$ being the radius of an unstable photon orbit.  Solving Eq.~(\ref{vr}) for Eq.~(\ref{rpot}) yields the critical impact parameters, \cite{Abdujabbarov:2016hnw,Kumar:2018ple}
\begin{align}
\xi_{c}=& \frac{(a^2+r^2)\Delta '(r)-4r \Delta (r) }{a \Delta '(r)} \nonumber\\
\eta_{c}=&\frac{r^2 \left(8 \Delta (r) \left(2 a^2+r \Delta '(r)\right)-r^2 \Delta '(r)^2-16 \Delta (r)^2\right)}{a^2 \Delta '(r)^2}\label{CriImpPara},
\end{align}.
These critical impact parameters separate the photon trajectories that fall into the black hole from those that escape to infinity. Hence, they are important for determining the boundary of the black hole shadow. The parameters $\xi_c$ and $\eta_c$ depend on the black hole spin parameter $a$ and the holonomy-correction parameter through the metric function $\Delta(r)$. As a result, the photon region and the shadow structure differ from those of the standard Kerr black hole spacetime.
The radii of circular photon orbits in the equatorial plane can be obtained by setting $\eta_c=0$, which corresponds to photon motion at $\theta=\pi/2$. The solutions of this equation give the radii of the prograde ($r_p^-$) and retrograde ($r_p^+$) photon orbits. These correspond to photons moving along and opposite to the direction of black hole rotation, respectively \cite{Bardeen:1973gs,Teo:2003ltt}. For the RHCBH spacetime, the equation takes the form
\begin{align}\label{etac=0}
64 \left(a^2-\left(b-r\right)\left(-2+r\right)\right)^2 r^2\notag\\
\Big(2a^2 - 16\left(a^2 - \left(b-r\right)\left(-2+r\right)\right)^2 \notag\\
+ r\left(-2 - b + 2r\right) - r^2\left(-2 - b + 2r\right)\Big)^2 = 0
\end{align}
\begin{figure*}
\centering
\begin{tabular}{c c }
 
    \includegraphics[scale=0.7]{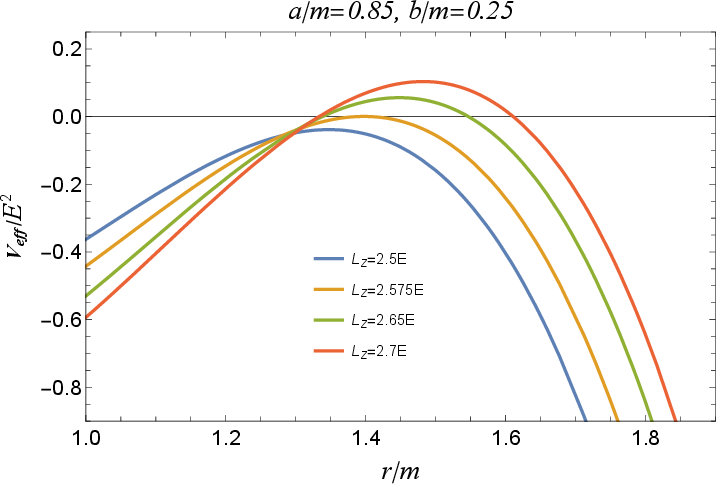}&
    \includegraphics[scale=0.7]{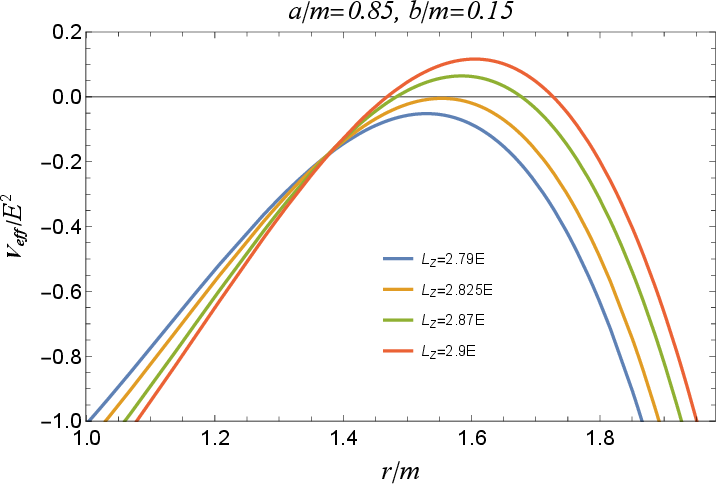}\\
    \end{tabular}

  \caption{Radial profile of the effective potential $V_{\text{eff}}/E^2$ for the RHCBH geometry as a function of the radial coordinate $r/m$. The plots shows the variation for distinct values of the axial angular momentum $L_z$ under fixed spacetime parameters: $a/m = 0.8$, $b/m = 0.25$ (left panel) and $a/m = 0.85$, $b/m = 0.15$ (right panel). The maxima of these curves determine the radii of the unstable circular photon orbits $r_p$ that delineate the shadow boundary .}
    \label{EFFECTIVEPOTENTIAL}
 \end{figure*}
\begin{figure*}
\centering
\begin{tabular}{c c }
 
    \includegraphics[scale=0.9]{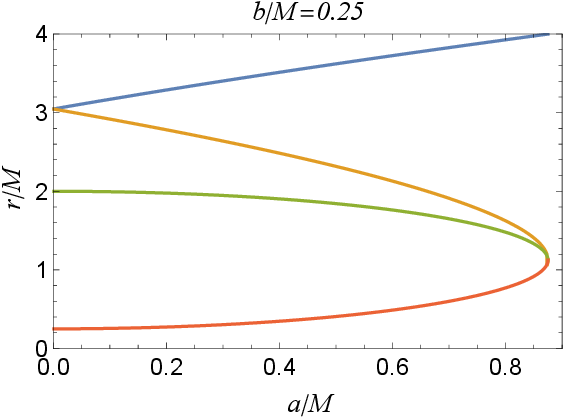}&
    \includegraphics[scale=0.9]{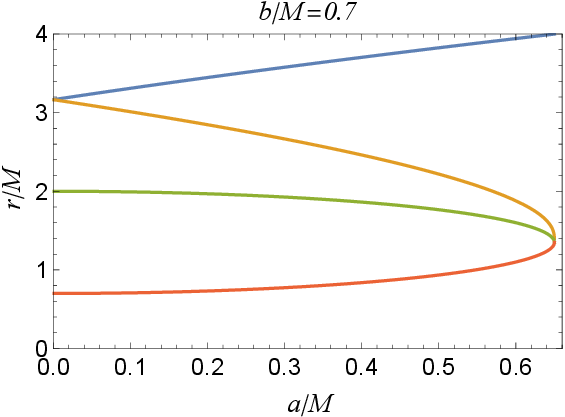}\\
    \end{tabular}

  \caption{Radii of various surfaces for the RHCBH as a function of spin $a$for a fixed $b$. The plot shows the inner Cauchy horizon $r_{-}$ (red), the outer event horizon $r_{+}$ (green), the prograde photon orbit $r_{p}^{-}$ (yellow), and the retrograde photon orbit $r_{p}^{+}$ (blue). The prograde orbit radius approaches the event horizon as the black hole approaches extremality.}
    \label{fig:proretro}
 \end{figure*}

The parameter $\eta_c$ determines the shape and nature of photon orbits around the black hole. When $\eta_c=0$, the photon motion remains confined to the equatorial plane. For $\eta_c>0$, photons can move along more general spherical or non-planar trajectories around the black hole \cite{Carter:1968rr,Chandrasekhar:1985kt,Teo:2003ltt}.
The motion of photons near the black hole depends on the values of it's impact parameters relative to the corresponding critical values. Photons with $\xi>\xi_c$ escape to spatial infinity, while those with $\xi<\xi_c$ are captured by the black hole. The special case $\xi=\xi_c$ corresponds to unstable circular photon orbits, which separate the escaping and captured trajectories. These unstable orbits form the boundary of the photon region and are responsible for the black hole shadow \cite{Bardeen:1973gs,Perlick:2021aok,Cunha:2018acu}.
Since the RHCBH spacetime is regular, captured photons do not end at a spacetime singularity. Instead, they spiral towards a central region with finite curvature. As these photons cannot return to a distant observer, they create the dark shadow region of the black hole \cite{Bardeen:1973gs,1972ApJ...173L.137C,Falcke:1999pj}. Photons moving along unstable spherical orbits form the bright photon rings around the shadow boundary \cite{Bardeen_1974,Johnson:2019ljv}.
In rotating black hole spacetimes, photons can move either along the direction of rotation or opposite to it. The orbits moving along the black hole rotation are called prograde orbits, while those moving opposite to the rotation are called retrograde orbits. Their corresponding radii in the equatorial plane are denoted by $r_p^{-}$ and $r_p^{+}$, respectively. These radii are obtained from the positive real roots of $\eta_c=0$ with the condition $r_p \geq r_+$, where $r_+$ is the event horizon radius. Due to the frame-dragging effect, the retrograde orbit always lies farther from the black hole than the prograde orbit \cite{Bardeen:1972fi,Johannsen:2010ru,Teo:2003ltt}.
The other spherical photon orbits lie in the range
 \begin{equation}
 r_p^- < r_p < r_p^+.
 \end{equation}
 For photons to cross the polar axis of the black hole, their axial angular momentum must vanish. The corresponding orbit radius, denoted by $r_p^0$, is obtained by setting $\xi_c=0$ and satisfies
 \begin{equation}
 r_p^- \leq r_p^0 \leq r_p^+.
 \end{equation}
Since the RHCBH metric depends on both the spin parameter $a$ and the holonomy-correction parameter $b$, the photon ring structure and the black hole shadow differ from those of the Kerr black hole. These changes modify the size and shape of the photon region and may produce observable signatures in black hole imaging observations such as the Event Horizon Telescope \cite{EventHorizonTelescope:2019dse,EventHorizonTelescope:2022wkp,Cunha:2018gql}.

In the limit $b=0$, Equation~(\ref{CriImpPara}) reduces to that of the standard Kerr black hole. In the equatorial plane of a Kerr black hole, the respective expressions for the prograde and retrograde circular photon orbits radii read \cite{Teo:2003}
 
\begin{eqnarray}\label{photonRKerr}
r_p^-&=&2M\left[1+ \cos\left(\frac{2}{3}\cos^{-1}\left[-\frac{|a|}{M}\right]\right) \right],\nonumber\\
r_p^+&=&2M\left[1+ \cos\left(\frac{2}{3}\cos^{-1}\left[\frac{|a|}{M}\right]\right) \right].
\end{eqnarray} 

\begin{figure*}
\centering
\begin{tabular}{c c }
 
    \includegraphics[scale=0.9]{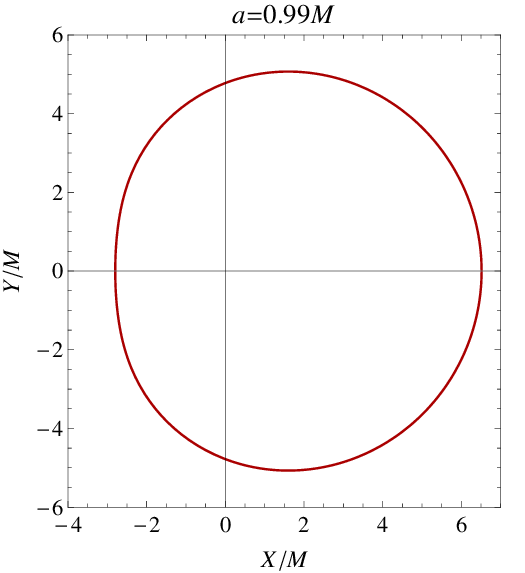}&
    \includegraphics[scale=0.9]{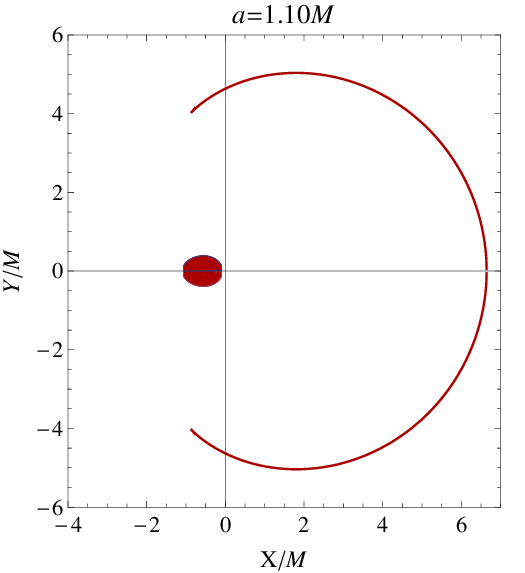}\\
    \end{tabular}

  \caption{A comparative analysis of shadow silhouettes cast by a standard Kerr black hole ($a \leq M$) with $a = 0.99M$ (left panel) and a Kerr naked singularity  ($a > M$) with $a = 1.10M$ (right panel) for an edge-on observer inclination ($\theta_o = 90^\circ$). The black hole exhibits a completely closed shadow boundary determined by a continuous photon region. In contrast, the naked singularity generates an unclosed, open arc structure due to the frame-dragging-induced destabilization and absence of a prograde photon ring.}
    \label{fig:KerrShadow}
 \end{figure*}
 \begin{figure}[hbt!]
\centering
\includegraphics[scale=0.85]{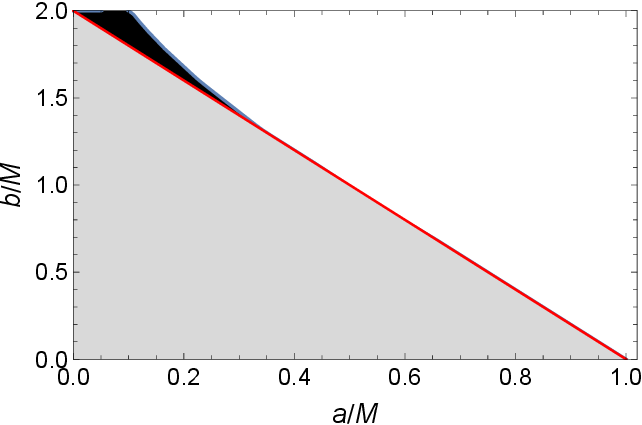}
\caption{Classification of the $(a, b)$ parameter space illustrating the transition between black hole horizons and horizonless geometries alongside their respective shadow. The grey region represents the black hole domain bounded by the extremal curve (red line). Within the exterior no-horizon domain, the black shaded region denotes the specific parameter window ($b_E \le b \le b_p$) where the spacetime supports unstable bound null geodesics, enabling the formation of a complete, closed photon ring despite the absence of an event horizon.} \label{fig:BHNoBH}
\end{figure}

The widening gap $\delta =  |r_P^- - r_E|$ with increasing $b$ can be understood as follows. The holonomy correction parameter $b$ modifies the radial dependence of the metric function
		\begin{equation}
			\Delta(r) = r^2 - (2M+b)r + a^2 + 2Mb.
		\end{equation}
		For fixed spin $a$, increasing $b$ weakens the effective gravitational potential near the center, as seen from the reduced magnitude of the second term in $\Delta(r)$. This reduction in the effective attractive force pushes the prograde photon orbits outward, increasing $r_P^-$. Simultaneously, the extremal horizon radius $r_E = M + b/2$ shifts outward as well, but more slowly than $r_P^-$ because $r_E$ depends linearly on $b$ while $r_P^-$ depends on $b$ through the critical impact parameters $\xi_c$ and $\eta_c$ which involve derivatives of $\Delta(r)$. Consequently, $\delta$ increases with $b$. This behavior is a direct manifestation of the LQG holonomy correction weakening the gravitational field in the strong-field region, consistent with the quantum regularization of the central singularity.

The radii obtained from the above expressions lie within the ranges $M \leq r_p^- \leq 3M$ and $3M \leq r_p^+ \leq 4M$. Here, $r_p^-$ and $r_p^+$ denote the radii of the prograde and retrograde photon orbits, respectively. Due to the rotation of the black hole, the prograde orbit lies closer to the event horizon, while the retrograde orbit is pushed farther away because of the frame-dragging effect \cite{Bardeen:1972fi,Chandrasekhar:1985kt}. In the Schwarzschild limit, corresponding to $a=0$, the spacetime becomes spherically symmetric and the distinction between prograde and retrograde motion disappears. In this case, Eq.~(\ref{etac=0}) describes the well-known photon sphere located at $r_p=3M$, which represents the unstable circular photon orbit around a Schwarzschild black hole \cite{Synge:1966okc,Claudel:2000yi}. For an extremal Kerr black hole with $a=M$, the prograde photon orbit approaches the event horizon and eventually coincides with it, giving $r_p^-=r_E=M$. This happens because of the strong dragging of spacetime near the rotating black hole \cite{Bardeen:1973gs,Teo:2003ltt}.

 \begingroup
\begin{table*}
	\caption{Prograde ($r_P^-$) and retrograde ($r_P^+$) photon orbit radii for the RHCBH at extremal parameter values. The table also shows the extremal horizon radius $r_E$ and the separation $\delta = |r_P^- - r_E|$ between the prograde photon orbit and the extremal horizon.}\label{table1A}

	\begin{ruledtabular}
		\begin{tabular}{l l l l l l}

$b_E/M$ & $a_E/M$ &$r_E/M$ & $r_{P}^{-}/M $ & $r_{P}^{+}/M $ & $\delta=| r_{P}^{-}-r_E| $\\
\hline \hline
 0 & 1 & 1 & 1 & 4 & 0\\
 0.25  & 0.875 & 1.125 & 1.125 & 4 &0  \\
 0.5  & 0.75 & 1.25  & 1.25  & 4 & 0\\
 0.75 & 0.625 & 1.375   & 1.5 & 4 &0.125 \\
 1  & 0.5 & 1.5 & 2 & 4 &0.5 \\
 1.25  & 0.375 & 1.625 & 2.5 &4 &0.875 \\
 1.5  & 0.25 & 1.75 & 3 & 4 &1.25\\
 1.75  & 0.125 & 1.875 & 3.5 & 4  &1.625\\
 2  & 0 & 2 & 4 & 4 &2  \\
 
		\end{tabular}
	\end{ruledtabular}
\end{table*}
\endgroup

Following Alonso-Bardaji {\it et al.} \cite{Alonso-Bardaji:2022ear}, we take the parameter $b$ in the form $b=2M\lambda^2/(\lambda^2+1)$. The quantity $\lambda$ is the dimensionless parameter associated with the polymerization of the effective LQG theory. Setting $\lambda=0$ gives $b=0$, and the corresponding metric reduces to the Schwarzschild case. The quantum-geometric input is usually expressed in terms of the area gap $\Delta=4\sqrt{3}\pi\gamma\ell_{\rm Pl}^2$. The relation between $\lambda$ and $\Delta$ is prescription dependent, so we treat $b$ as the parameter that controls the quantum correction in the spacetime considered here \cite{Ashtekar:2011ni, Ashtekar:1996eg}.

Next, we relate the celestial coordinates $(X,Y)$ on the image plane of the observer and the critical impact parameters $\xi_c$ and $\eta_c$. Using the geodesic equations (\ref{tuch}), (\ref{th}), and (\ref{phiuch}) along with the tetrad components of the four-momentum $p^{(\mu)}$, an observer can measure the apparent position of the photon located at ($r_o,\theta_o$) as:

\begin{eqnarray}
&&X= -r_o\frac{p^{(\phi)}}{p^{(t)}} = -\left. r_o\frac{\xi_c}{\sqrt{g_{\phi\phi}}(\zeta-\gamma\xi_c)}\right|_{(r_o,\theta_o)},\nonumber\\
&&Y = r_o\frac{p^{(\theta)}}{p^{(t)}} =\pm\left. r_o\frac{\sqrt{\mathcal{V}_{\theta}(\theta)}
}{\sqrt{g_{\theta\theta}}(\zeta-\gamma\xi_c)}\right|_{(r_o,\theta_o)},~\label{Celestial}
\end{eqnarray} 
where
\begin{eqnarray}
\zeta=\sqrt{\frac{g_{\phi\phi}}{g_{t\phi}^2-g_{tt}g_{\phi\phi}}},\qquad \gamma=-\frac{g_{t\phi}}{g_{\phi\phi}}\zeta.
\end{eqnarray}

The off-diagonal metric component $g_{t\phi}$, which appears due to the rotation of the black hole, together with the angular potential $\mathcal{V}_{\theta}(\theta)$, determines the values of the celestial coordinates $X$ and $Y$. These coordinates describe the apparent position of photons in the observer's image plane. Here, $X$ measures the displacement perpendicular to the projected axis of rotation of the black hole, while $Y$ measures the displacement parallel to the rotation axis \cite{Bardeen:1973gs,Chandrasekhar:1985kt,Perlick:2021aok}. 
For an observer located far away from the black hole in an asymptotically flat region $(r_o\to\infty)$, the celestial coordinates simplify to the well-known expressions \cite{Bardeen:1973gs,Hioki:2009na}:

\begin{equation}
X=-\xi_c\csc\theta_o,\qquad Y=\pm\sqrt{\eta_c+a^2\cos^2\theta_o-\xi_c^2\cot^2\theta_o}.\label{pt}
\end{equation} 

The above equation for an observer located in an equatorial plane ($\theta_o=\pi/2$) simplifies to
\begin{equation}
X=-\xi_c,\qquad Y=\pm\sqrt{\eta_c}.\label{pt1}
\end{equation}

\begin{figure*}
\begin{tabular}{c c }
	\includegraphics[scale=0.7]{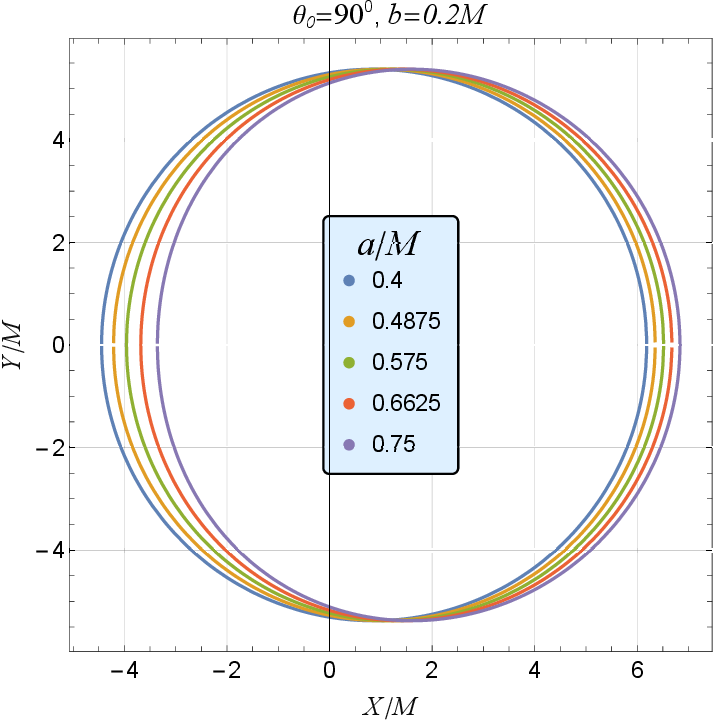}&
	\includegraphics[scale=0.7]{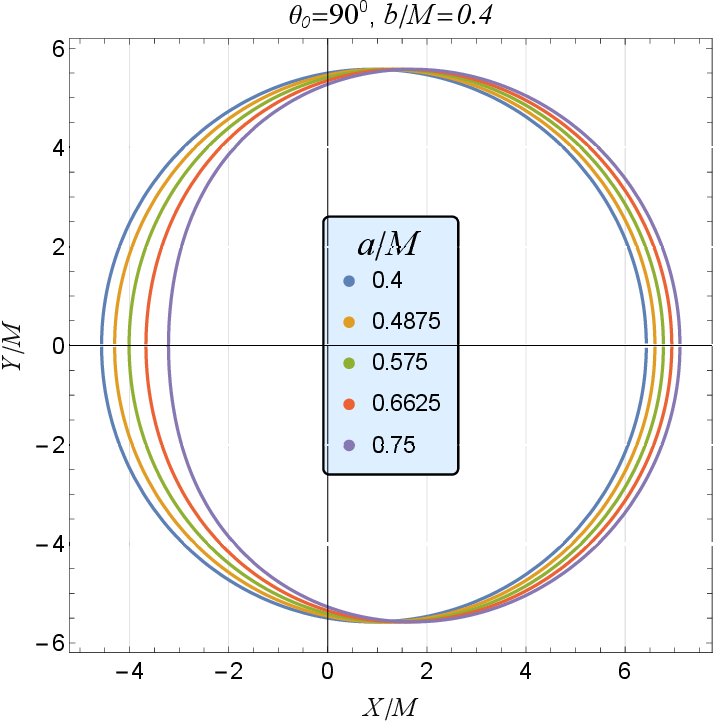}\\

\end{tabular}
	\caption{Shadow silhouettes in the celestial image plane $(X/M, Y/M)$ for the RHCBH at an inclination of $\theta_o = 90^\circ$, demonstrating the impact of the spin $a/M$ and the quantum correction parameter $b$. The left panel ($b = 0.2M$) and right panel ($b = 0.4M$) illustrate that increasing $b$ causes an overall enlargement of the shadow silhouette, while increasing the spin parameter $a/M$ induces asymmetry characterized by a vertical flattening of the prograde edge and a horizontal shift of the shadow center. }\label{fig:BHShadow}
\end{figure*}

\begin{figure*}
\centering
  \includegraphics[scale=0.65]{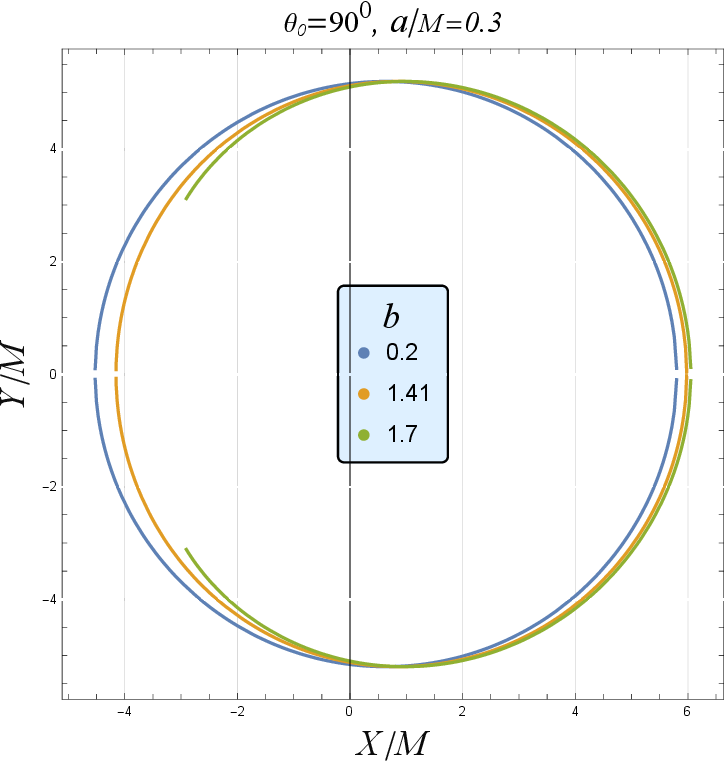}
  \caption{Comparative shadow silhouettes at a viewing angle of $\theta_o = 90^\circ$ and fixed spin $a/M = 0.4$ detailing the transition from black hole to no-horizon spacetime configurations. The blue curve represents a standard black hole shadow ($b = 0.2 \le b_E$). In the no-horizon regime ($b > b_E$), the yellow curve ($b = 1.41$) demonstrates that a fully closed shadow boundary is maintained within the window $b_E \le b \le b_p$. In contrast, the green curve ($b = 1.7$) reveals an open arc-like structure once $b$ exceeds the threshold value $b_p$.  }\label{fig:NoBHShadow}
\end{figure*}

The geometry of the black hole shadow is determined by the choice of its mass function $m(r)$, hence the parametric plots generated in the $(X,Y)$ plane using Eqs.~(\ref{Celestial}) or (\ref{pt}), yield a wide variety of black hole shadows \cite{Abdujabbarov:2016hnw,Amir:2016cen,Kumar:2019pjp,Kumar:2018ple,Kumar:2020ltt}. The simplest case of $m(r)=M$ reproduces the shadows in Kerr geometry \cite{Bardeen1973,Kumar:2018ple}.

For viewing angles $\theta_o=0$ and $\theta_o=\pi$, the shadow appears perfectly circular for all values of the spin parameter $a$ because of the axial symmetry of the spacetime \cite{Bardeen:1973gs,Hioki:2009na}. Kerr black holes with $a\leq M$ possess both prograde and retrograde unstable photon orbits. These orbits together form a closed photon ring, which determines the boundary of the black hole shadow \cite{Kumar:2020yem,Kumar:2020ltt,Perlick:2021aok}. The situation is different for a Kerr naked singularity with $a>M$. In this case, photons moving in prograde equatorial orbits, as well as some nearby spherical photon orbits, are strongly affected by frame dragging and fall directly into the singularity instead of forming closed paths. However, photons in retrograde orbits are less affected by the rotation and can still escape to a distant observer. As a result, the observed image changes from a closed shadow ring to an arc-like structure with a missing segment near the left side of the image plane \cite{Hioki:2009na,Charbulak:2018wzb,Stuchlik:2019uvf}. Photons with orbit radii satisfying $r_p<0$ also play an important role when $\theta_o\neq \pi/2$. These photons move towards another asymptotically flat region corresponding to $r\to -\infty$ and therefore do not reach the observer located at $r_o\to\infty$ \cite{Hioki:2009na,DeFalco:2021ksd}. The absence of these photon trajectories produces the dark central region seen in the shadow image (cf. Fig.~\ref{fig:KerrShadow}).
The two different shadows, closed rings for RHCBHs vs. open arcs for Kerr naked singularities, provide a clear observational test to distinguish between these two cases. If future high-resolution VLBI observations were to reveal an open arc-like structure, it would favor Kerr naked singularities over RHCBHs. Conversely, the fact that RHCBHs produce closed shadows even without horizons demonstrates that the \textit{lack} of an open arc does not necessarily imply the existence of an event horizon.

The Cosmic Censorship Conjecture (CCC) asserts that spacetime singularities are always hidden by event horizons, so that they cannot be seen by external observers \cite{Penrose:1969pc}. The validity of the conjecture remains an open question, as it has not been proven to date despite its foundational role in GR. In order to be consistent with the proposition of CCC, the presence of an event horizon guarantees a closed photon ring in Kerr spacetime. Interestingly, in certain alternative spacetime geometries, such as the spherically symmetric Janis-Newman-Winicour \cite{Shaikh:2019hbm} and Joshi-Malafarina-Narayan \cite{Shaikh:2018lcc} spacetimes, naked singularities admit closed photon spheres for some ranges of their parameter values and consequently cast shadows closely resembling those of Schwarzschild black holes. In contrast, in the absence of photon spheres, the naked singularity images are markedly different from those of black holes \cite{Shaikh:2019hbm}. This demonstrates the fact that, rather than the mere presence of an event horizon, the existence of a photon sphere determines whether the shadow formed is closed or not. In the context of our RHCBH model, we find the scenario particularly intriguing; the regular core structure induced by the holonomy correction parameter $b$ resolves curvature singularities even in parameter ranges where event horizons do not exist. 

Figure~\ref{fig:BHShadow} shows the shadow silhouettes of the RHCBH obtained by plotting the celestial coordinates $(X,Y)$ in terms of the photon orbit radius $r_p$. In both panels, the shadow boundaries form closed curves, showing that the RHCBH spacetime supports unstable spherical photon orbits for the chosen parameter values, consistent with the discussion in the previous paragraph.
As the spin parameter $a$ increases, the shadow centre shifts horizontally and the boundary on the prograde side becomes more flattened. This distortion is caused by the rotation of the black hole and the associated frame-dragging effect \cite{Bardeen:1973gs,Hioki:2009na}.  We also observe that increasing the holonomy-correction parameter $b$ increases the overall size of the shadow, especially its width along the $X$-axis. This suggests that the parameter $b$ weakens the gravitational field near the central region and pushes the photon sphere outward. As a result, photons can orbit the black hole at larger radii, leading to a larger shadow structure \cite{Cunha:2015yba,Konoplya:2019sns}.
\subsection{Shadows of horizonless RHCBH spacetimes}
In our analysis, we find that no-horizon RHCBH spacetimes can still produce a closed shadow boundary for parameter values in the range $b_E<b\leq b_p$ (cf. Figs.~\ref{fig:BHNoBH}, \ref{fig:NoBHShadow}). This behaviour is different from the Kerr naked singularity, where the shadow generally changes into an open arc-like structure when the event horizon disappears \cite{Kumar:2020ltt,Hioki:2009na}. The existence of a closed shadow in the absence of an event horizon shows that the shadow structure is mainly governed by the presence of unstable photon orbits rather than the horizon itself.
We further observe that the critical parameter $b_p$, which separates closed shadow rings from open arc-like structures, depends on the inclination angle $\theta_o$. This shows that the shadow structure is affected not only by the spacetime geometry but also by the observer's viewing direction. In other words, changing the inclination angle can modify the observed shape of the shadow from a closed ring to an open arc-like pattern \cite{Cunha:2018acu,Perlick:2021aok}.
These results may have important observational implications. The high-resolution shadow images of M87$^\ast$ and Sgr A$^\ast$ obtained by the Event Horizon Telescope (EHT) \cite{EventHorizonTelescope:2019dse,EventHorizonTelescope:2022wkp} provide a possible way to test quantum-corrected rotating spacetimes. In particular, the differences between the shadow structures of RHCBH/no-horizon geometries and those of Kerr black holes or Kerr naked singularities may help in identifying possible deviations from classical general relativity \cite{Cunha:2015yba,Konoplya:2019sns}.

The comparison highlights the key conceptual insight that closed shadows do not require event horizons. The RHCBH provides an explicit counterexample to the common assumption that shadows imply horizons. This is a valuable contribution to the ongoing debate on distinguishing black holes from horizonless compact objects \cite{Kumar:2020ltt, Shaikh:2019hbm, Cunha:2018gql}.

\subsection{Discussion on the Cosmic Censorship Conjecture}
The Cosmic Censorship Conjecture (CCC), proposed by Penrose \cite{Penrose:1969pc} states that spacetime singularities produced by gravitational collapse should be hidden behind event horizons and should not be visible to distant observers. In the RHCBH spacetime, the central curvature singularity is removed by holonomy corrections and replaced by a regular minimal spacelike hypersurface at $r=b$. For $b_E<b\leq b_p$, the spacetime has no event horizon but it remains regular, with finite curvature everywhere.
The absence of an event horizon does not imply a violation of the CCC our case. The CCC is mainly concerned with whether a singularity can be seen by a distant observer. Here, the central singularity is removed by the holonomy corrections, and the spacetime is regular everywhere. Hence, there is no singularity that needs to be hidden by an event horizon. The horizonless nature of the RHCBH solution therefore does not provide a direct counterexample to the CCC.

Based on our results, we extend the discussion of the CCC to regular spacetimes. A regular object without an event horizon can still behave like a black hole if it has unstable circular photon orbits that form a closed shadow. This happens for the RHCBH solution in the range $b_E<b\leq b_p$. Although there is no event horizon in this range, the photon region gives a complete closed shadow ring. The resulting shadow can be very similar to the shadow of a classical black hole and may therefore be difficult to distinguish using shadow observations alone. If the required photon orbits are absent, the observed structure can instead be open or arc-like, which may be distinguishable with future high-resolution observations.
Thus, for regular horizonless spacetimes, the discussion can be shifted from ``Does an event horizon exist?'' to ``Does the spacetime possess the photon region required to produce a closed shadow?'' This provides a useful observational perspective on the CCC and on the possibility of regular black hole mimickers. The same point is relevant for interpreting present and future EHT and ngEHT observations, and is consistent with earlier studies of photon rings and horizonless compact objects \cite{Shaikh:2019hbm, Kumar:2020ltt, Kumar:2022fqo,Cunha:2018gql}.

\begin{figure*}
\centering
	\includegraphics[scale=0.75]{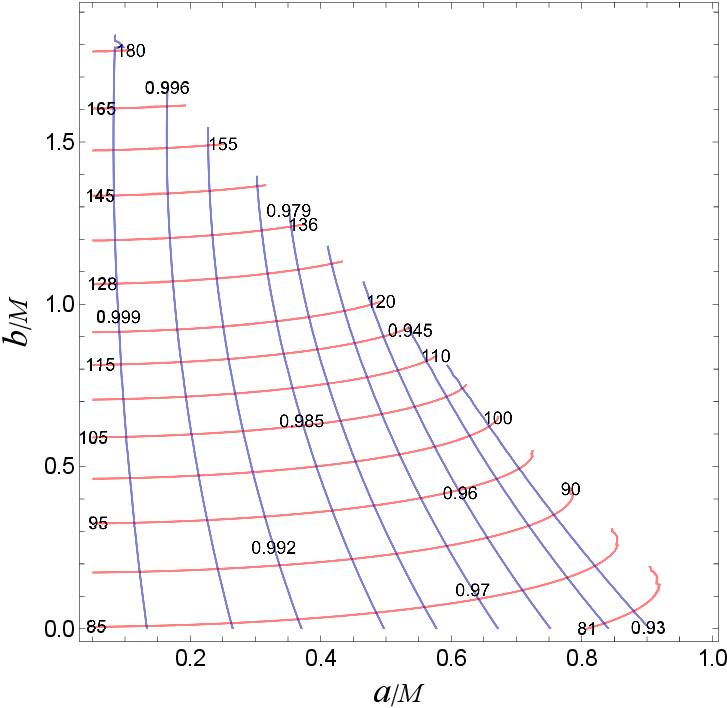}
	
	\caption{Contour plots of the shadow observables in the RHCBH parameter space $(a/M, b/M)$ for an equatorial observer. The red lines map curves of constant shadow area $A$, and the blue lines map contours of constant shadow oblateness $D$. The unique discrete intersection points of these two independent sets of contours simultaneously determine the spin $a$ and holonomy parameter $b$.}\label{PE}
\end{figure*}

\begin{table*}[hbt!]
\caption{Estimated values of RHCBH parameters $b/M$ and $a/M$ derived from shadow observables using the Kumar-Ghosh method. The parameters are determined from the intersection points of constant shadow area $A$ and oblateness $D$ contours for an equatorial observer ($\theta=90^\circ$). }\label{parameter_table1}
\centering
\begin{tabular}{|l|l|l|l| }
\hline
$A$ & $D$ & $a$/M & $b/M $\\
\hline \hline
85 & 0.996 & 0.261 & 0.01751 \\ 
\hline
85 & 0.992 & 0.3637 & 0.03381 \\
\hline
85 & 0.985 & 0.4827 & 0.05123 \\
\hline
85 & 0.96 & 0.7081 & 0.1231 \\ 
\hline
95 & 0.999 & 0.1126 & 0.3324 \\
\hline
95 & 0.992 & 0.3136 & 0.3446 \\
\hline
95 & 0.979 & 0.4806 & 0.3814\\
\hline
95 & 0.945 & 0.6704 & 0.4663 \\ 
\hline
105 & 0.996 & 0.2005 & 0.6014 \\
\hline
105 & 0.992 & 0.2796 & 0.6125 \\
\hline
105 & 0.979 & 0.429 & 0.6419 \\ 
\hline
105 & 0.96 & 0.542 & 0.6874 \\
\hline
115 & 0.999 & 0.09092 & 0.8179 \\
\hline
115 & 0.992 & 0.2576 & 0.8364\\
\hline
115 & 0.979 & 0.3953 & 0.8625 \\ 
\hline
128 & 0.996 & 0.1724 & 1.071 \\
\hline
128 & 0.992 & 0.24 & 1.08 \\
\hline
128 & 0.97 & 0.417 & 1.125 \\ 
\hline
145 & 0.996 & 0.1658 & 1.342 \\
\hline
145 & 0.992 & 0.2298 & 1.349 \\
\hline

\end{tabular}

\end{table*}

\section{Shadow observables and parameter estimation: Kumar-Ghosh Method}\label{sect4}
Estimating the physical parameters of black holes has been an important topic in gravitational physics for several decades. Early studies mainly relied on the motion of stars and gas around compact objects to determine black hole masses and spins \cite{Bardeen:1970zz,Thorne:1974ve}. Later, techniques based on X-ray spectroscopy, continuum fitting, quasi-periodic oscillations, and gravitational-wave observations were developed to probe the properties of astrophysical black holes \cite{Reynolds:2013qqa,LIGOScientific:2016aoc}. The possibility of directly imaging black hole shadows has opened a new way to study strong gravity near the event horizon \cite{Falcke:1999pj}. In particular, the Event Horizon Telescope (EHT) observations of M87$^\ast$ and Sgr A$^\ast$ have provided the first direct measurements of shadow structures and offered new opportunities to test general relativity in the strong-field regime \cite{EventHorizonTelescope:2019dse,EventHorizonTelescope:2022wkp}.
The shadow of a rotating compact object carries important information about the spacetime geometry and can therefore be used to estimate the physical parameters of the source \cite{Bardeen:1973gs,Hioki:2009na,Cunha:2018acu}. Over the years, several shadow observables have been introduced to describe the size, shape, and distortion of black hole shadows and to distinguish between different gravitational models \cite{Tsupko:2017rdo,Kumar:2020yem}. 

The possibility of estimating black hole parameters from shadow observations received significant attention after the work of Falcke {\it et al.} \cite{Falcke:1999pj}, who showed that the shadow of a supermassive black hole could be observed using very long baseline interferometry. Later, Psaltis {\it et al.} \cite{Psaltis:2014mca} and Johannsen  {\it et al.} \cite{Johannsen:2010ru,Johannsen:2015hib} demonstrated that accurate measurements of the shadow size and shape can provide important information about black hole parameters such as mass and spin, and can also be used to test possible deviations from the Kerr geometry.  Hioki and Maeda method \cite{Hioki:2009na} uses the shadow observables $R_s$ and $\delta_s$ to estimate black hole parameters from the black hole shadow. Later, Tsupko \cite{Tsupko:2017rdo} extended this method and derived analytical expressions for parameter estimation. Another related approach was developed by Tsukamoto \textit{et al.} \cite{Tsukamoto:2014tja} to distinguish Kerr black hole shadows from those arising in modified theories of gravity. These methods mainly assume that the shadow boundary is nearly circular and possesses certain symmetries. Such assumptions work reasonably well for Kerr black holes with moderate spin \cite{Bardeen:1973gs,Hioki:2009na}. However, in many modified gravity models and non-Kerr rotating spacetimes, the shadow boundary can become highly distorted and asymmetric \cite{Schee:2008kz,Johannsen:2015qca,Abdujabbarov:2015xqa,Younsi:2016azx}. In addition, realistic observational data may contain noise and finite-resolution effects, which can further deform the observed shadow shape \cite{Falcke:1999pj,Psaltis:2014mca,Abdujabbarov:2015xqa}. In such cases, observables based solely on circular symmetry may not yield accurate parameter estimates. Motivated by these limitations, Kumar and Ghosh \cite{Kumar:2018ple} introduced a new set of observables that can describe more general and distorted shadow shapes. Since Kumar - Ghosh method does not depend on circular symmetry, it is better suited for studying rotating non-Kerr black holes, modified gravity spacetimes, and quantum-corrected geometries where the shadow may significantly deviate from a circular shape \cite{Kumar:2020yem,Shaikh:2019hbm}.  This makes it particularly suitable for studying rotating non-Kerr black holes, regular spacetimes, and no-horizon geometries. In this work, we use the Kumar--Ghosh method \cite{Kumar:2018ple}, which employs suitable geometric observables constructed from the shadow boundary to estimate the black hole spin and additional spacetime parameters \cite{Kumar:2020yem,Kumar:2020ltt}.  Since the RHCBH spacetime depends on both the spin parameter $a$ and the holonomy-correction parameter $b$, the corresponding shadow observables may help constrain possible quantum-corrected effects using future high-resolution black hole shadow observations. 

\subsection{Shadow area and oblateness observables}
Kumar and Ghosh \cite{Kumar:2018ple} proposed a more reliable set of shadow observables based on the shadow area, $A$ and the oblateness $D$. These observables are applicable even for the non--circular haphazard shadow silhouettes. These are defined by:

\begin{eqnarray}
A &=& 2\int{Y(r_p) \, dX(r_p)} = 2\int_{r_p^{-}}^{r_p^+} \left( Y(r_p) \frac{dX(r_p)}{dr_p} \right) dr_p, \label{eq:Area}
\end{eqnarray}
\begin{eqnarray}
D &=& \frac{X_r - X_l}{Y_t - Y_b}, \label{eq:Oblateness}
\end{eqnarray}

where $(X_l, X_r)$ and $(Y_b, Y_t)$ denote the leftmost, rightmost, bottommost, and topmost extremal coordinates of the shadow boundary in the image plane.

The oblateness parameter $D$, for an equatorial observer ($\theta=\pi/2$), serves as a natural measure of the shadow's deviation from perfect circularity. In the non-rotating limit ($a=0$), $D=1$ for a perfectly circular shadow -- the result of a Schwarzschild photon sphere. With increase in spin, the shadow gets deformed due to frame--dragging resulting in decrease in the value of $D$, which reaches its minimum value of $D = \sqrt{3}/2$ for the extremal Kerr black hole ($a=M$) \cite{Tsupko:2017rdo}. The dependence of $A$ and $D$ on both the parameters $a$ and $b$ makes these observables a helpful tool in the estimation of parameters of RHCBHs. 

\begin{figure*}

\begin{tabular}{c c}
	\includegraphics[scale=0.75]{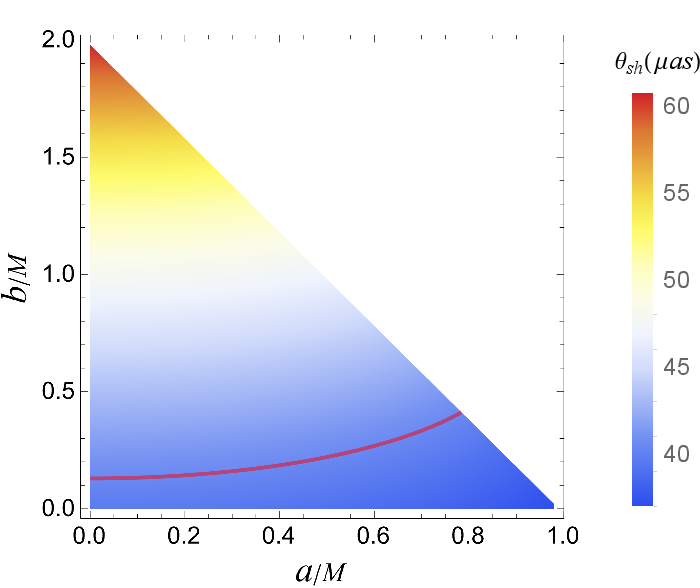}&
	\includegraphics[scale=0.75]{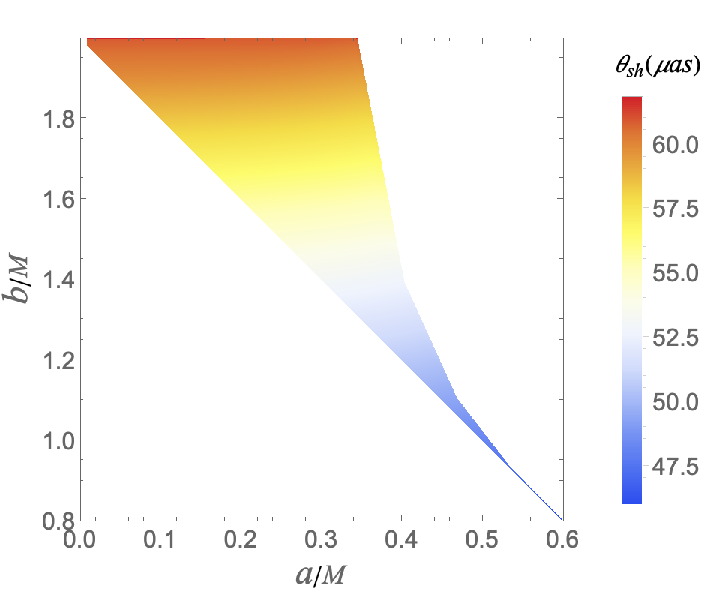}
\end{tabular}

	\caption{ (Left): RHCBH shadow angular diameter $\theta_{\text{sh}}$ (in $\mu\text{as}$) spanning the parameter space $(a/M, b/M)$ for a representative $\text{M87}^*$ line-of-sight inclination angle of $\theta_o = 17^\circ$. The left panel details the black hole domain where the solid red line marks the $\theta_{\text{sh}} = 40.5\,\mu\text{as}$ upper bound, enclosing the parameter combinations consistent with the $1\sigma$ EHT observation. The right panel provides the corresponding angular diameter distribution across the regular no-horizon spacetime.\label{fig:M8717}}
    \end{figure*}

\begin{figure*}
        \begin{tabular}{c c}
	\includegraphics[scale=0.75]{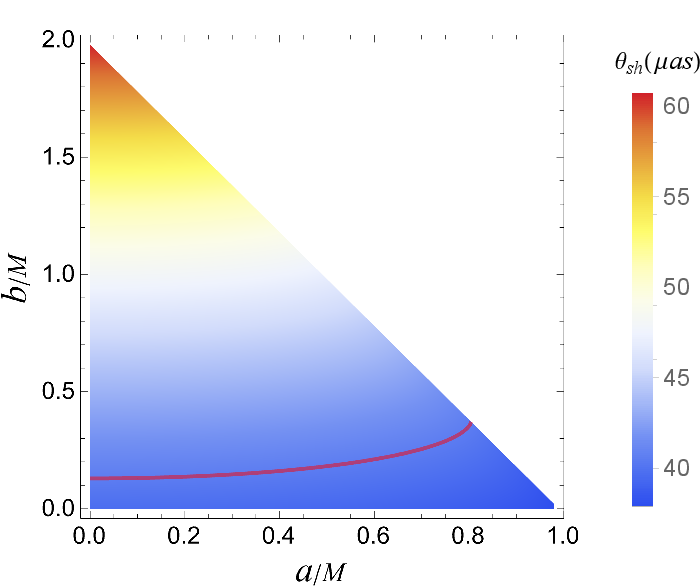}&
	\includegraphics[scale=0.75]{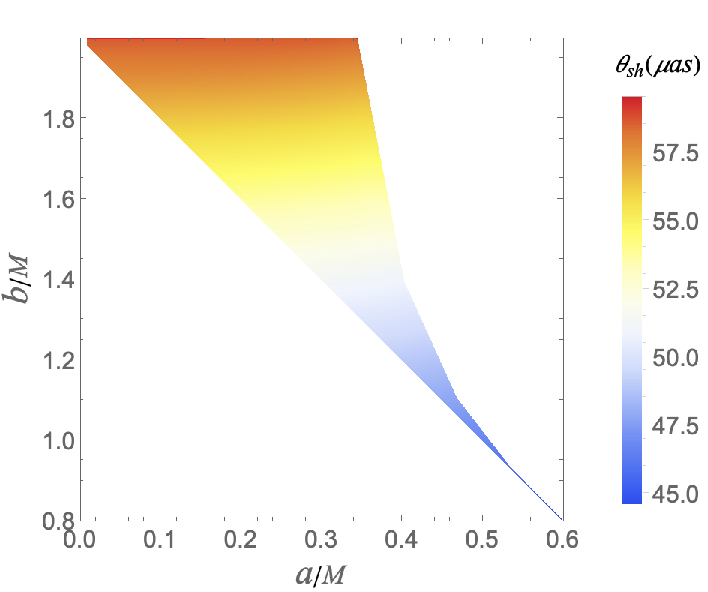}	
\end{tabular}	
	\caption{ RHCBH shadow angular diameter $\theta_{\text{sh}}$ (in $\mu\text{as}$) evaluated at an edge-on inclination angle of $\theta_o = 90^\circ$ where rotational distortion effects are maximized. The solid red line in the black hole phase space (left panel) corresponds to the $\theta_{\text{sh}} = 40.5\,\mu\text{as}$ threshold matching the $\text{M87}^*$ size constraints. The right panel shows the shadow angular-diameter values for the counterpart regular no-horizon configurations. \label{fig:M8790}}
\end{figure*}

\begin{figure*}[ht]
\begin{tabular}{c c}
	\includegraphics[scale=0.75]{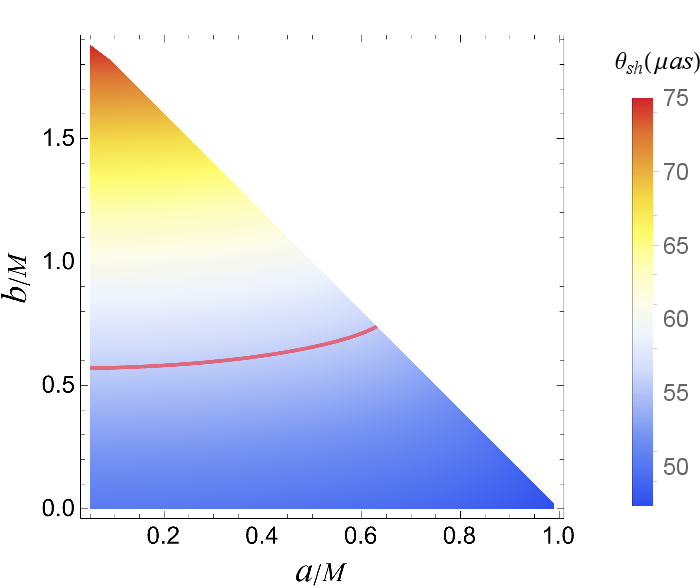}&
	\includegraphics[scale=0.75]{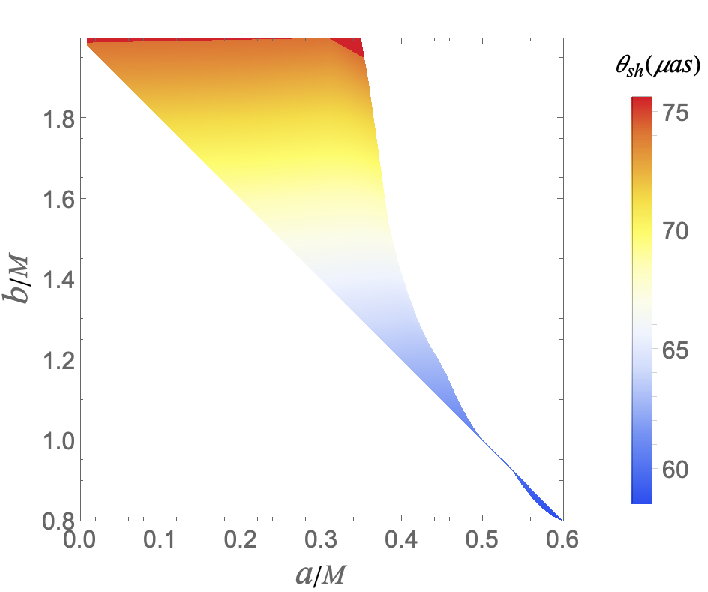}
\end{tabular}	
	\caption{ (Left:) RHCBH shadow angular diameter $\theta_{sh}$ (in $\mu\text{as}$) spanning the parameter space $(a/M, b/M)$ for a representative SgrA*  line-of-sight inclination angle of $\theta_o = 50^\circ$. The left panel details the black hole domain where the solid red line marks the $\theta_{\text{sh}} = 55.7\,\mu\text{as}$ upper bound, enclosing the parameter combinations consistent with the $1\sigma$ EHT observation. The right panel provides the corresponding angular diameter distribution across the regular no-horizon spacetime.\label{fig:SGR50}}
\end{figure*}

\begin{figure*}[ht!]
\begin{tabular}{c c}
	\includegraphics[scale=0.75]{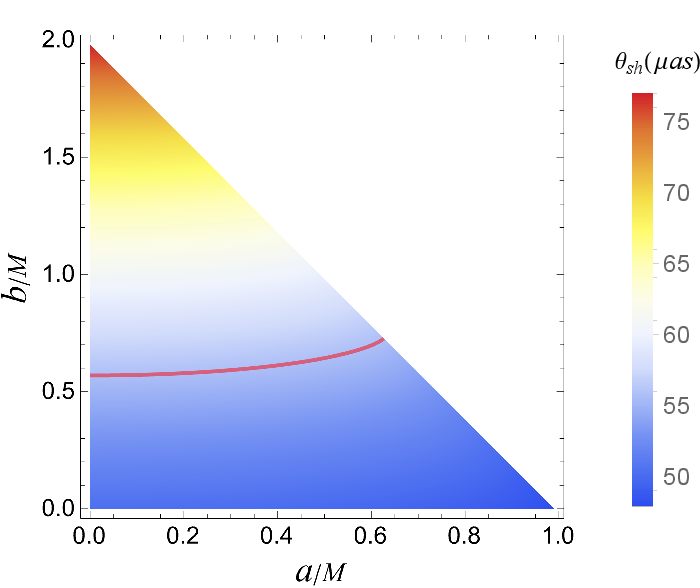}&
	\includegraphics[scale=0.75]{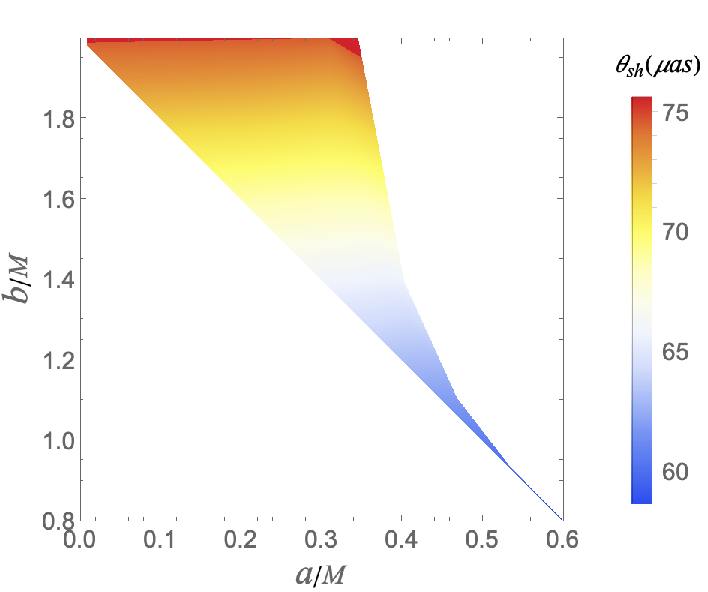}
\end{tabular}	
	\caption{ RHCBH shadow angular diameter $\theta_{\text{sh}}$ (in $\mu\text{as}$) evaluated at an edge-on inclination angle of $\theta_o = 90^\circ$ where rotational distortion effects are maximized. The solid red line in the black hole phase space (left panel) corresponds to the $\theta_{\text{sh}} = 55,7\,\mu\text{as}$ threshold matching the $\text{SgrA}^*$ size constraints. The right panel shows the shadow angular-diameter values for the counterpart regular no-horizon configurations. \label{SGR90}}
    \end{figure*}

\begin{figure*}
     \includegraphics[scale=0.75]{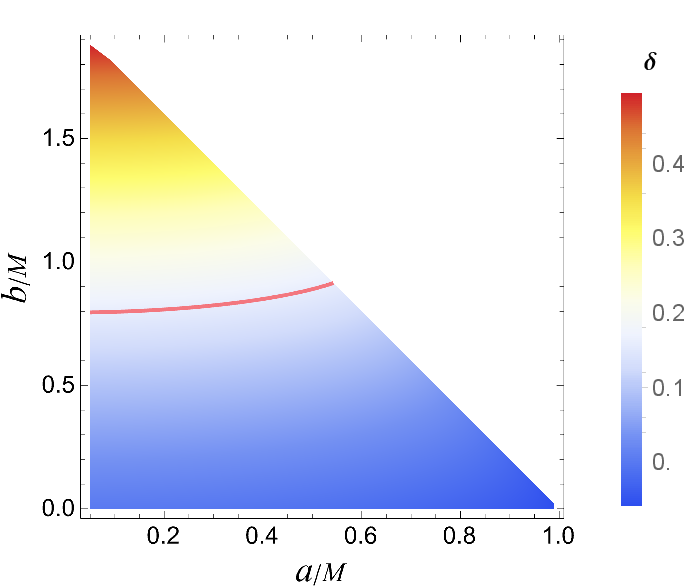}
	\includegraphics[scale=0.75]{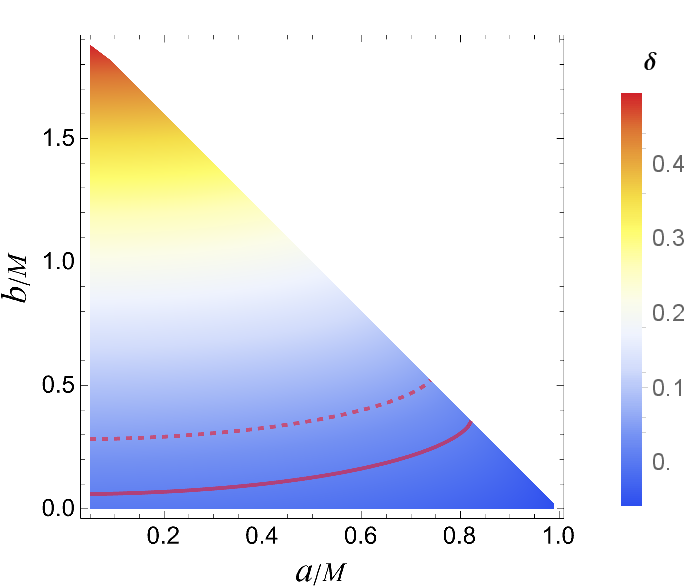}

	\caption{RHCBH shadow angular diameter deviation from that of a Schwarzschild black hole as a function of ($a, b$). The solid red line corresponds to the upper bound of the EHT inferred value of Schwarzschild deviation of M87* black hole (left panel), $\delta_{M87^*}=-0.01\pm0.17$. The dashed and solid red lines correspond to the upper bounds of EHT inferred values for Sgr A* (right panel): $-0.04_{-0.10}^{+0.09}$ (Keck), $\delta_{SgrA^*}=-0.08_{-0.09}^{+0.09}$ (VLTI). }\label{Deviation}
\end{figure*}

In parameter estimation of black holes, we often encounter the challenge of resolving degeneracies introduced by modified gravity parameters. Here, in addition to the modified gravity parameter, the parameter $a$ and inclination angle $\theta_{0}$ are responsible for degeneracies. The shadow gets enlarged and distorted by both -- $a$ and $b$, while the inclination angle $\theta_{0}$ also determines the degree of asymmetry \cite{Bardeen:1973tla,Luminet:1979nyg,Hioki:2009na}. Hence, the parameters $a$ and $b$ can produce qualitatively identical shadows when varied independently, complicating the extraction of parameters using observations. The Kumar-Ghosh method \cite{Kumar:2018ple} used here helps to break the degeneracy between the spin $a$ and the holonomy parameter $b$ by considering two shadow observables, the area $A$ and the oblateness $D$. As shown in Fig.~\ref{PE}, the constant $A$ and constant $D$ contours cross at a definite point in the $(a,b)$ plane, which gives the corresponding values of $a$ and $b$. Using only one of these observables would not provide the same constraint. We also calculate the shadow angular diameter $\theta_{\rm sh}$ and the Schwarzschild deviation parameter $\delta$ throughout the allowed $(a,b)$ region, as shown in Figs.~\ref{fig:M8717}, \ref{fig:M8790}, \ref{fig:SGR50}, \ref{SGR90}, and \ref{Deviation} rather than considering only a few selected parameter values. The $1\sigma$ bounds are then obtained using the uncertainties reported by the EHT.

The shadow observables $A$ and $D$ have been widely used to test models against EHT observations. For example, ~\cite{Kumar:2022vfg} applied the Kumar--Ghosh method to Loop Quantum Gravity motivated multihorizon rotating black holes. Similarly, Islam et al.~\cite{Islam:2022wck} employed the same framework to constrain LQG-inspired rotating black holes. More recently, Ali, Islam, and Ghosh~\cite{Ali:2024ssf} used the shadow area  and oblateness  to constrain rotating quantum-corrected black holes using EHT observations of both M87* and Sgr A*. Here, we apply this method to the RHCBH metric, a novel LQG-motivated holonomy-corrected geometry. To the best of our knowledge, this is the first study to obtain quantitative constraints on the holonomy parameter  of the RHCBH spacetime using the shadows of both M87* and Sgr A* by applying an established observational method to a new black hole geometry.

\section{Constraints from EHT observations}\label{sect5}

The first horizon-scale images of supermassive black holes (SMBHs) M87* and Sgr A* were released by the Event Horizon Telescope (EHT) collaboration in 2019 \cite{EventHorizonTelescope:2019dse,EventHorizonTelescope:2019ggy} and 2022 \cite{EventHorizonTelescope:2022wkp,EventHorizonTelescope:2022xqj}, respectively. These observations provided the first direct images of black hole shadow structures and opened a new avenue for testing gravity in the strong-field regime. The observed shadow 
 and angular size were found to be broadly consistent with the predictions of the Kerr black hole in general relativity \cite{Psaltis:2007cw,Johannsen:2010ru,Hioki:2009na}. At the same time, the EHT observations also provide an important opportunity to test possible deviations from the Kerr geometry and to constrain alternative theories of gravity and quantum-corrected black hole models \cite{Bambi:2019tjh,Vagnozzi:2022moj,Afrin:2021imp}. The shadow properties of a black hole are determined by the underlying photon region and therefore carry direct information about the spacetime geometry \cite{Falcke:1999pj,Cunha:2018acu,Perlick:2021aok}. Various shadow observables, such as the shadow size, distortion, and angular diameter, can be used to extract the physical parameters of the black hole and to distinguish between different gravitational theories \cite{Hioki:2009na,Tsupko:2017rdo,Kumar:2018ple}. Here, we use the EHT measurements of the shadow angular diameter together with the bounds on the Schwarzschild deviation parameter $\delta$ to constrain the holonomy-correction parameter $b$ and the spin parameter $a$ of the RHCBH spacetime.

Using the expression for the shadow area $A$ from Eq.~(\ref{eq:Area}), the areal radius of the shadow is defined as \cite{Kumar:2018ple}
\begin{eqnarray}
R_a=\sqrt{A/\pi}.
\label{Arealradius}
\end{eqnarray}
The corresponding angular diameter of the black hole shadow is given by \cite{Bambi:2019tjh,Kumar:2020owy,Afrin:2021imp}
\begin{eqnarray}
\theta_{sh}=2\frac{R_a}{d},
\label{angularDiameterEq}
\end{eqnarray}
where $d$ is the distance between the black hole and the observer. The angular diameter connects the theoretically calculated shadow size with the EHT observations.
To quantify the deviation of the shadow from the Schwarzschild case, we use the Schwarzschild deviation parameter introduced by the EHT collaboration \cite{EventHorizonTelescope:2022urf,EventHorizonTelescope:2022xqj}:
\begin{equation}
\delta=\frac{\theta_{sh}}{\theta_{sh,\rm Sch}}-1.
\end{equation}
where $\theta_{sh, Sch}$ corresponds to the Schwarzschild diameter. A positive (negative) value of  $\delta$ indicates that the shadow radius is larger (smaller) than that of a Schwarzschild black hole with the same mass. 

In the Kerr black hole, with variation of $a/M \in [0,1]$ and $\theta_{0} \in[0,90^\circ]$ can result in a at the most $7.5 \%$ shrinkage of the shadow diameter relative to the Schwarzschild diameter corresponding to $-0.075\leq\delta\leq0$, thereby in agreement with the EHT bounds. Deviation outside this interval signals a departure from Kerr geometry \cite{EventHorizonTelescope:2022xqj}. Since the shadow size depends on the photon region governed by the holonomy correction parameter $b$, the EHT bounds on $\delta$ constrain the allowed region of the parameter space $(a,b)$.

\subsection{Constraints from the EHT results of M87*}
The first horizon-scale image of M87*, obtained by the Event Horizon Telescope (EHT) using very-long-baseline interferometry (VLBI) observations at 230 GHz \cite{EventHorizonTelescope:2019dse,EventHorizonTelescope:2019ths,EventHorizonTelescope:2019ggy}, revealed a bright, nearly circular crescent-shaped emission ring surrounding a central dark region. This dark region is interpreted as the black hole shadow produced by strong gravitational lensing and relativistic effects near the photon region of the black hole \cite{EventHorizonTelescope:2019dse}. The measured angular diameter of the emission ring is
\begin{align*}
    \theta_d = 42 \pm 3\,\mu{\rm as},
\end{align*}
where the uncertainty mainly arises from observational and instrumental limitations \cite{EventHorizonTelescope:2019dse}. 
The interpretation of the observed ring is, however, not completely straightforward because of the sparse VLBI baseline coverage and uncertainties associated with accretion-flow modelling, plasma emission, and radiative transfer effects \cite{Gralla:2020pra,Gralla:2019xty,Chael:2021rjo}. In addition, the exact physical origin of the observed ring is still under discussion. While a significant part of the emission is expected to arise from strongly lensed photons near the photon ring, contributions from the surrounding accreting plasma and emission structure may also play an important role \cite{Gralla:2020pra,Gralla:2019xty,Johnson:2019ljv}.

Taking into account a possible $\leq 10 \%$ offset between the observed photon ring and the actual photon ring, the calibrated shadow angular diameter of M87* is estimated as \cite{EventHorizonTelescope:2019dse, Afrin:2024khy}:
\begin{align*}
    \theta_{sh}=37.8\pm2.7\mu as,
\end{align*}
where the error of $\pm2.7\%$ incorporates both  the potential offset and the measurement error. We adopt the EHT inferred values of mass and distance from earth -- $M_{M87^*} = 6.5\times10^9 M_\odot$ and $d_{M87^*} = 16.8 Mpc$ to impose bounds on the parameters $a$ and $b$ using $\theta_{sh}$ and $\delta$. The relativistic jet oriented at $163$\textdegree~ relative to the line of sight  \cite{CraigWalker:2018vam}, in the symmetric shadow analysis, is equivalent to $17$\textdegree~, which encourages us to depict the angular diameter of the RHCBH shadow of M87* for $\theta_o=17$\textdegree~. We also consider $\theta_o=90$\textdegree~ since the shadow distortion is maximum at this angle. We obtain the corresponding bounds on the parameter space ($a/$, $b/M$) within the $1\sigma$ range, $[35.1-40.5]\mu$as and to establish bounds on ($a/M$, $b/M$), we use $\theta_{sh}=40.5 \mu as$ . The M87* black hole shadow angular diameter places a bound on the parameters ($a$, $b$) where $b \in (0, 0.1319 M^{-1})$ at $a=0$, $b \in (0, 0.421 M^{-1})$ at $a=0.784 M$, for $\theta_o=17$\textdegree~ (cf. Fig.~\ref{fig:M8717}) (left) and $b \in (0, 0.1316 M^{-1})$ at $a=0$ , $b \in (0, 0.3685 M^{-1})$ at $a=0.803 M$ for $\theta_o=90$\textdegree~ (cf. Fig.~\ref{fig:M8790}) (left).

Another constraint comes from the Schwarzschild deviation parameter inferred by the EHT Collaboration. By calibrating the shadow size with the observed emission ring diameter, the EHT analysis obtained \cite{EventHorizonTelescope:2021dqv,Afrin:2024khy}
$\delta_{M87^*}=-0.01\pm0.17 $ at the $1\sigma$ confidence level.  If the RHCBH shadow to satisfy this observational bound further restricts the allowed region in the $(a/M,b/M)$ parameter space.
Using the observational limits on $\delta$, we obtain the bounds
$
b/M \in (0,0.8028)
\quad \text{at} \quad a=0,
$
and
$
b/M \in (0,0.919)
\quad \text{at} \quad a=0.5355\,M,
$
for the observer inclination angle $\theta_o=90^\circ$ [cf. Fig.~\ref{Deviation}(left)]. These results show that nonzero values of the holonomy correction parameter remain consistent with current EHT observations across a substantial region of parameter space. This suggests that the RHCBH geometry can successfully reproduce the observed shadow of M87* without conflicting with the present observational uncertainties \cite{EventHorizonTelescope:2019ggy,EventHorizonTelescope:2021dqv}.

\subsection{Observational constraints from the EHT results of Sgr A*}
The 2017 VLBI mission of EHT, at 1.3mm wavelength, led to the first ever horizon scale image of the Sgr A* black hole \cite{EventHorizonTelescope:2022exc,EventHorizonTelescope:2022urf,EventHorizonTelescope:2022apq,EventHorizonTelescope:2022wkp, EventHorizonTelescope:2022wok,EventHorizonTelescope:2022xqj}. Owing to its smaller angular size and rapid variability, imaging SgrA* is significantly more challenging compared to M87*. However, the images of its shadow are particularly important for strong gravity tests, since they probe curvature scales approximately $10^{6}$ higher than those of M87*, while its mass-to-distance ratio is independently determined from previous estimates.

Independent stellar dynamic observations of the observations of the S0-2 star with the Keck telescopes and the Very Large Telescope Interferometer (VLTI) \cite{Do:2019txf,Chael:2021rjo,Nicolas:2023cnc,EventHorizonTelescope:2022xqj} led to the determination of the mass and distance of Sgr A*  $M = 4.0_{-0.6}^{+1.1} \times 10^6 M\odot$ and $d = 8$ kpc \cite{EventHorizonTelescope:2022wkp,EventHorizonTelescope:2022xqj}. The reported angular diameter of the shadow is $d_{sh} = 48.7 \pm 7,\mu$as. We determine $\theta_{sh}$, which, in addition to other black hole parameters, is based on the mass $M$, the distance $d$ of the black hole, the parameter $b$, and the inclination $\theta_0$ as depicted in Figs.~\ref{fig:SGR50} (left) and \ref{SGR90} (left). The EHT collaboration, using the calibration method, has obtained a shadow angular diameter of $ SgrA^* = 48.7 {\pm} 7\mu as$. We shall use  $d_{sh}^{Sgr A*}$ as 55.7 $\mu as$. The need for theoretical predictions to fall within the EHT uncertainty range yields bounds on the parameters $a$ and $b$.

In Figs.~\ref{fig:SGR50} (left) and \ref{SGR90} (left), we calculate the angular diameter of the RHCBH shadow of SgrA* for $\theta_o=50$\textdegree~ and 90\textdegree~  as a function of ($a/M$, $b/M$). We used $\theta_{sh}=55.7 \mu as$ to put the bounds, hence constraining the parameters  ($a$, $b$) such that $b \in (0, 0.5764 M^{-1})$ at $a=0$ , $b \in (0, 0.7482 M^{-1})$ at $a=0.6253 M$ for $\theta_o=50$\textdegree~ and $b \in (0, 0.5738 M^{-1})$ at $a=0$ , $b \in (0, 0.7241 M^{-1})$ at $a=0.6239 M$  for $\theta_o=90$\textdegree~ (cf. Figs.~\ref{fig:SGR50} (left) and \ref{SGR90} (left)).  The agreement of a considerable part of the plots of $\theta_{sh}$ with the EHT observations indicates that the moderate deviations from Kerr induced by the parameter $b$ are not excluded by the present data of SgrA*.

An additional constraint is provided by the Schwarzschild deviation parameter $\delta$. Using separate angular size priors from the Keck and Very Large Telescope Interferometer (VLTI) and the three independent imaging models, EHT estimated the bounds on the fraction deviation observable $\delta$ \cite{EventHorizonTelescope:2022exc, EventHorizonTelescope:2022xqj}
\begin{align}
\delta= \begin{dcases*} -0.08^{+0.09}_{-0.09}\;\;\;\;\; & \text{VLTI}\\
		  -0.04^{+0.09}_{-0.10}\;\;\;\;\; &\text{Keck}	 
		\end{dcases*} 
\end{align}
in agreement with the Kerr expectation ($-0.075\leq  \delta \leq 0$).
We set the bounds on the RHCBH parameters and impose the inferred limitations of EHT  on $\delta$. We report that modeling Sgr A* as an RHCBH sets bounds on its parameters for inferred bound Keck (-0.14,0.05) and VLTI  (-0.17,0.01) observations. The bounds obtained are as (cf. Fig.~\ref{Deviation}) (right):
\begin{enumerate}
    \renewcommand{\labelenumi}{\roman{enumi}.}
    \item  For \textbf{Keck}:
    At $a=0.04856$, $b \in (0, 0.2854 M^{-1})$  , at $a=0.7348 M$, $b \in (0, 0.5141 M^{-1})$ . 
    \item For \textbf{VLTI}:
    At $a=0.05334$, $b \in (0, 0.06087 M^{-1})$  , at $a=0.8176 M$, $b \in (0, 0.358 M^{-1})$ .
\end{enumerate}

Strong evidence that the RHCBHs can be viable candidates for another supermassive black hole is provided by the parameter space's good consistency with the observational results for Sgr A* (cf. Fig.~\ref{Deviation}) (right). The Sgr A* results provide an independent test, complementing those of M87*, since the curvature scale of the latter is vastly different from that of the former. 
\section{Comparison with other rotating black hole spacetimes} \label{sect6}
In this section, we compare the constrained parameters of RHCBHs with those of several well-motivated rotating black holes that possess additional deviation parameters beyond Kerr including rotating quantum corrected black holes (RQCBHs) \cite{Ali:2024ssf}, LQG-inspired rotating black holes (LIRBHs) \cite{Afrin:2022ztr}, rotating black holes in Bumblebee gravity \cite{Islam:2024sph},  hairy Kiselev black holes \cite{Ahmed:2025zdc} and four widely studied rotating regular black holes namely Bardeen \cite{Bambi:2013ufa,Kumar:2020yem,Kumar:2020ltt}, Hayward \cite{Bambi:2013ufa,Kumar:2020yem,Kumar:2020ltt}, Ghosh \cite{Ghosh:2014pba} and Simpson-Visser black holes \cite{Mazza:2021rgq,Islam:2021ful}, whose spacetimes are free of curvature singularities everywhere \cite{KumarWalia:2022aop}. Each of the spacetimes mentioned above can be expressed in the general form of the metric \ref{rotmetric} with an appropriately modified mass function $M(r)$, thereby enabling us to compare their shadow predictions with the EHT--constrained values (cf. Table~\ref{bhconst}).

\begin{table}[h!]
\caption{Constraints on deviation parameters for various rotating black hole spacetimes from EHT observations of M87* and Sgr A* within the $1\sigma$ confidence region. The table compares constraints from modified theories of gravity (LQG-inspired, Bumblebee gravity, hairy Kiselev) with four well-known regular black hole models: Bardeen, Hayward, Ghosh, Simpson-Visser. Parameters $a$, $\ell$, $l$, and $g$ are in units of $M$, while $\alpha$ is in units of $M^2$.}
\label{tab:comparative_constraints }
\centering
\begin{tabular}{l l l}
\hline
\multirow{2}{*}{\textbf{Rotating black hole metric in modified theories of gravity}} &  \multirow{2}{*}{\textbf{Constraints from M87*}} & \multirow{2}{*}{\textbf{Constraints from SgrA*}}\\
& &   \\

\hline
\multirow{2}{*}{Ali-Ghosh QCBHs  \cite{Ali:2024ssf}} &$0.6157\leq a\leq0.8511$  & $0<a<0.8066$ \\
&  $0 < \alpha< 0.8985$ & $0 < \alpha< 1.443$  \\
\hline
\multirow{2}{*}{ Hairy Kiselev black holes \cite{Ahmed:2025zdc}}  & $0.2402 < l < 1 $ & $0.03304< l < 0.7415 $  \\
& for $\omega=-2/3$ & for $\omega=-2/3$   \\
\hline
\multirow{2}{*}{Bumblebee gravity \cite{Islam:2024sph}}  &  $-0.0037< \ell < 0.1672  $  & $-0.1162< \ell <0.01123 $   \\
& for $a=0.5$ & for $a=0.5$\\
\hline
\multirow{2}{*}{LIRBH-1  \cite{Afrin:2022ztr}}& \multirow{2}{*}{$0 < L_q< 0.0643$} &\multirow{2}{*} {$0 < L_q< 0.0423$} \\
&  &    \\
\hline
\multirow{2}{*}{LIRBH-2  \cite{Afrin:2022ztr}}& \multirow{2}{*}{ $0 < L_q< 0.1253$} & \multirow{2}{*}{$0 < L_q< 0.0821$}   \\
&   &   \\
\hline
\multirow{2}{*}{Ghosh black holes
 \cite{KumarWalia:2022aop}} & $0\leq g\leq0.30461$ & $ 0.155 \leq g \leq 0.61116.$ \\
& for $a=0.7$ & for $a=0.2$ \\ 
\multirow{2}{*}{Hayward blackholes \cite{KumarWalia:2022aop}} &$ 0\leq g\leq0.73627$ & $g\geq 0.4881$\\
& for $a=0.7$   & for $a=0.2$\\
\hline 
\multirow{2}{*}{Bardeen black holes \cite{KumarWalia:2022aop}} & $ 0\leq g\leq0.30182$ &$0.154 \leq g \leq 0.58367$  \\
& for $a=0.7$ & for  $a=0.2$\\
\hline
\hline 
\multirow{2}{*}{Simpson–Visser black holes \cite{KumarWalia:2022aop}}  & \multirow{2}{*}{independent of $g$} & \multirow{2}{*}{independent of $g$} \\
& & \\
\hline

\end{tabular}

\label{bhconst}
\end{table}

\paragraph{Ali-Ghosh quantum corrected black holes}
Ali and Ghosh \cite{Ali:2024ssf} obtained the rotating quantum-corrected black holes (AGQCBHs) spacetime from the spherically symmetric seed metric obtained by  Lewandowski {\it et al.} \cite{Lewandowski:2022zce} using MNJA \cite{Azreg-Ainou:2014pra}. AGQCBHs resolve the classical singularity within the LQG framework. The resulting spacetime can be described by the metric \ref{rotmetric} with the mass function 
\begin{equation}
M(r)=  M - \frac{\alpha M^2}{2r^{3}},
\end{equation}
with $\alpha$ being the quantum correction parameter encoding the effects of LQG. Using the EHT shadow angular diameter, the parameters are constrained to $a \in (0.6157,\, 0.8511)\,M$ and $0 < \alpha < 0.8985\, M^2$ for M87*, and $a \in (0,\, 0.8066)\,M$ and $0 < \alpha < 1.443\, M^2$ for Sgr A* (cf. Table~\ref{bhconst}). Unlike the linear dependence of $\Delta(r)$ on $b$ in RHCBH, $\alpha$ appears as a quadratic term in the mass function of RQCBH. Therefore, the shadow observables exhibit a distinct dependence on the quantum-correction parameter.

\paragraph{Hairy Kiselev black holes}
Ahmed {\it et al.} \cite{Ahmed:2025zdc} used the gravitational decoupling approach along with the MNJA \cite{Brahma:2020eos,Yang:2022yvq} to obtain the rotating hairy Kiselev black hole solution. This spacetime is characterized by a hair parameter $l$ and a quintessence parameter $\omega$, and can be described by metric (\ref{rotmetric}), with the corresponding mass function given by:
\begin{equation} 
M(r) = M-\alpha\frac{r}{2} e^{-r/(M-l/2)}+\frac{1}{2}\frac{1}{r^{3\omega+1}}.
\end{equation}
For  $\omega = -2/3$, using the EHT observations for M87* and Sgr A*, the hair parameter is restricted to $0.2402M < l< 1$  and  $0.03304M < l< 0.7415M$ respectively \cite{Ahmed:2025zdc} (cf. Table~\ref{bhconst}). While in RHCBH, the parameter $b$ modifies $\Delta(r)$ linearly, the mass function of Hairy Kiselev Black Holes exhibits an exponential dependence on $l$ leading to a qualitatively different sensitivity of the shadow observables to the deviation parameter. 

\paragraph{Bumblebee gravity black holes}
Islam {\it et al.} \cite{Islam:2024sph} explored this model, in which Lorentz violation, a vector-tensor extension of Einstein-Maxwell theory, is parametrised by $\ell$. A vector field, whose vacuum expectation value is non-vanishing, is introduced by this model, thereby inducing spontaneous Lorentz symmetry breaking \cite{Kostelecky:1989jw,Bluhm:2004ep}. The rotating black hole solution is described by metric (\ref{rotmetric}) with mass function
\begin{equation}
    M(r) = \frac{M(1+\frac{r\ell}{2M})}{1+\ell}
\end{equation}
Islam {\it et al.} \cite{Islam:2024sph} constrained the parameters of this model using shadow angular diameter of M87* and Sgr A*. For $a = 0.5M$, the M87* bounds yield $-0.0037M \leq \ell \leq  0.1672M$, while Sgr A* provides the tighter constraint $-0.1162M$ to $ 0.01123M$ (cf. Table~\ref{bhconst}). In contrast to holonomy correction parameter $b$, $\ell$ can assume both positive and negative values, reflecting the distinct nature of the Lorentz--violating correction.

\paragraph{LQG-inspired rotating black holes(LIRBHs)}
LIRBHs introduce a deviation parameter $L_q$ that encodes quantum effects. Afrin {\it et al.} \cite{Afrin:2022ztr} investigated two such models: LIRBH-1, which is basically the rotating counterpart \cite{Liu:2020ola} of the semiclassical spherically symmetric solution \cite{Modesto:2008im}, and LIRBH-2, derived using MNJA \cite{Brahma:2020eos,Yang:2022yvq} from the quantum--extended Schwarzschild metric \cite{Bodendorfer:2019nvy} acting as a seed metric. Using the EHT-inferred values for the shadow diameters of M87* and Sgr A* constrains the parameter $L_q$ to $0 < L_q< 0.0643$ and $0 < L_q< 0.0423$ for LIRBH-1 and $0 < L_q< 0.1253$ and $0 < L_q< 0.0821$ respectively, as summarized in Table~\ref{bhconst}.

\paragraph{Ghosh black holes}
\cite{Ghosh:2014pba} represent a distinct class of black holes with an asymptotically Minkowski core \cite{Simpson:2019mud}, unlike the de Sitter cores of Bardeen and Hayward regular black holes. This metric found by Ghosh \cite{Ghosh:2014pba} generalizes the spherically symmetric regular black hole solution \cite{Culetu:2014lca} to the rotating case, with mass function as:
\begin{eqnarray}
M(r)=Me^{-g^2/2Mr},
\end{eqnarray}
where $g$ is the NED charge.
The EHT constraints yield $0\leq g \leq 0.30461M$ for $a=0.7M$ from M87* \cite{Kumar:2020yem} and $g$ as $0.155M<g\leq 0.61116M$ for $a=0.2M$ \cite{Kumar:2020ltt} from SgrA*.  

\paragraph{ Hayward black holes}
Hayward (\citeyear{Hayward:2005gi}) presented this spacetime with an additional length--scale parameter $\ell$, governing the region concentrating the central energy density in a way that modifications to the metric become significant when the curvature scale approaches $\ell^{-2}$. This model provides an exact solution to a framework of NED minimally coupled to general relativity with magnetic charge $g$ is related to $\ell$ through $g^3=2M\ell^2$ \cite{Fan:2016hvf}. The rotating Hayward spacetime can be described by the metric (\ref{rotmetric}) with the mass function as: \cite{Hayward:2005gi,Kumar:2018ple}
\begin{equation}
M(r)=\frac{Mr^3}{r^3+g^3}\label{Haywardmass}
\end{equation}
Imposing M87* shadow angular diameter of $\theta_{sh} = 39.6192\; \mu$as within the $1 \sigma$ bound of EHT constrains $g$ to $0<g \leq 0.73627M$ for $a=0.7M$ \cite{Kumar:2020yem}, while the SgrA* observation within $46.9\,\mu\text{as} \leq \theta_{sh} \leq 50\,\mu\text{as}$ yields $g\geq0.4881M$ for $a=0.2M$ \cite{Kumar:2020ltt}. 

\paragraph{ Bardeen black holes}
The rotating Bardeen spacetime \cite{Bambi:2013ufa,Kumar:2020ltt,Kumar:2018ple,Kumar:2020yem,Ghosh:2015pra} can be defined by metric (\ref{rotmetric}) with the mass function \cite{Bardeen:1968} as:
\begin{equation}
M(r)=M\left(\frac{r^2}{r^2 + g^2}\right)^{3/2}, \label{Bardeenmass}
\end{equation}
where the magnetic monopole charge, g, \cite{AyonBeato:2000zs} $g$ is the deviation parameter. The shadow of the rotating Bardeen black hole ($g\neq0$) shrinks and becomes more distorted with increasing $g$, relative to that of the Kerr Black Hole\cite{KumarWalia:2022aop,Abdujabbarov:2016hnw,Kumar:2020yem,Kumar:2020ltt,Tsukamoto:2014tja}. The shadow angular diameters of M87* and Sgr A* constrain the parameter $g$ to $ 0\leq g\leq0.30182M$ for $a=0.7M$ and  $0.154M \leq g \leq 0.58367M$ for  $a=0.2M$ respectively \cite{Kumar:2020ltt,Kumar:2020yem}. 

\paragraph{ Simpson-Visser black holes}
The rotating Simpson-Visser black hole metric is obtained as a modification of the Kerr black hole metric \cite{Shaikh:2021yux,Mazza:2021rgq}, and can be described by (\ref{rotmetric}) with the mass function as:
\begin{eqnarray}
M(r)=M \sqrt{1+ \frac{g^2}{r^2}}.
\end{eqnarray}
In these black holes, $\Sigma = r^2+g^2 +a^2\cos^2\theta$ and  $\Delta = r^2+g^2+a^2-2 M(r) r$ differ from those of the three previously mentioned regular black holes. A distinctive feature of this spacetime is that the photon sphere, the areal radius and the shadow do not depend on $g$ with the proper distance between these surfaces being $g-$dependent \cite{Lima:2021las}.

The comparison presented in this section provides a distinct framework for testing quantum gravity through holonomy corrections. The different spacetimes, each characterised by a different deviation parameter, highlight the challenges in constraining non-Kerr parameters, since each case yields shadow predictions consistent with EHT measurements at current precision. Higher-precision data are required to break this degeneracy, to place sharper bounds on non-Kerr deviations and identify the true nature of astrophysical black holes more precisely.

\section{Conclusion}\label{sect7}
After decades of thinking about quantum gravity only as a theory for the Planck scales of the universe, we can now finally test it. The EHT has given us images of black hole shadows. This opens up a real possibility: maybe we can see quantum imprints on spacetime itself. Connecting LQG with real observations has always been a huge challenge. The EHT data now gives us a genuine chance to test whether holonomy-corrected black holes match what is observed in the Universe. Motivated by the above, we have conducted a systematic investigation of RHCBHs from LQG, focusing on their shadows and observational viability in light of Event Horizon Telescope data for M87* and Sgr A*.  Our analysis leads to several key conclusions -- some are useful for future observations, others help us think differently about quantum gravity. We list them below.
\begin{enumerate}
 \item We used the Hamilton-Jacobi formalism to understand the motion of photons in this spacetime. The analytically derived critical impact parameters--$\xi_c$ and $\eta_c$,  determine whether a photon gets captured, scattered, or moves in an unstable orbit. The retrograde and prograde orbits shift outward relative to their Kerr counterparts upon incorporation of the holonomy correction parameter $b$. The position of the retrograde orbit is independent of $b$ and remains fixed at $r_p^+ = 4M$ while the position of the prograde orbit depends on $b$, resulting in an outward shift with increasing $b$.  This happens because the quantum correction weakens the gravity near the center, making the shadow larger.
\item An increase in the parameter $b$ at a fixed spin results in an enlarged shadow and an increase in the spin at a fixed $b$ results in the horizontal shift of the shadow centroid, flattening the prograde edge, consistent with the frame-dragging effect. 
\item  When $b$ becomes large enough so that the event horizon of black holes disappears, the spacetime still remains regular without any singularity forms. However, unlike ordinary Kerr naked singularities, which give open arc-like shadows, these horizonless RHCBHs can still produce complete, closed shadow rings as long as $b$ is not too large (cf. Fig.~\ref{fig:BHNoBH}) within the range $b_E \leq b \leq b_p$. It tells us that closed shadows come from unstable photon orbits, not from the horizon itself. So if we ever see a closed shadow, we cannot automatically say a horizon exists. The critical boundary that separates closed rings from open arcs explicitly depends on the inclination $\theta_o$ of the observer.
\item We adopted the Kumar--Ghosh method to estimate the parameters using the shadow area $A$ and oblateness $D$ as the two primary observables. The intersection of constant $A$ and $D$ contours in the $(a, b)$ plane at unique points allows the determination of both parameters simultaneously. The results summarized in Table~\ref{parameter_table1} imply that the observables $A$ and $D$, when taken together, resolve the degeneracy which neither of the observables would have been able to do on their own.
This gives us a practical way to measure quantum gravity parameters from future, sharper images of black hole shadows.
\item For M87$^{*}$, the angular diameter bound $\theta_{\mathrm{sh}}$ constrains  $b \in (0, 0.1319 M^{-1})$ at $a=0$, $b \in (0, 0.421 M^{-1})$ at $a=0.784 M$, for $\theta_o=17$\textdegree~ (cf. Fig.~\ref{fig:M8717}) (left) and  $b \in (0, 0.1316 M^{-1})$ at $a=0$ , $b \in (0, 0.3685 M^{-1})$ at $a=0.803 M$ for $\theta_o=90$\textdegree~(cf. Fig.~\ref{fig:M8790}) (left), while the Schwarzschild deviation $\delta_{M87^*} = -0.01 \pm 0.17$ allows comparatively wider range $b \lesssim 0.919\,M$ at $a = 0.5355\,M$. For Sgr~A$^{*}$, the angular diameter bound constrains $b \in (0, 0.5764 M^{-1})$ at $a=0$ , $b \in (0, 0.7482 M^{-1})$ at $a=0.6253 M$ for $\theta_o=50$\textdegree~ and $b \in (0, 0.5738 M^{-1})$ at $a=0$ , $b \in (0, 0.7241 M^{-1})$ at $a=0.6239 M$  for $\theta_o=90$\textdegree~ (cf. Figs.~\ref{fig:SGR50} (left) and \ref{SGR90} (left)). 
\item  The Keck and VLTI Schwarzschild deviation priors yield individual constraints at $b \lesssim 0.5141\,M$ at $a = 0.7348\,M$ and $b \lesssim 0.358\,M$ at $a = 0.8176\,M$, respectively with the VLTI prior providing the tightest constraint on $b$ among all observables considered. This agreement of the RHCBH geometry with both EHT datasets across a substantial region of parameter space is a meaningful result confirming that the nonzero holonomy corrections are not excluded by current EHT data.
\item We compared RHCBHs with several other modified gravity and regular black hole models: quantum-corrected black holes, LQG-inspired models, Bumblebee gravity, hairy black holes, and regular ones like Bardeen, Hayward, Ghosh, and Simpson-Visser. All of them remain consistent with current EHT data. The way $\Delta(r)$ depends on $b$ in our model is different (linear) from many others (quadratic or exponential), but the present observations are not sharp enough to tell these apart. This means the models are not physically the same: our telescopes just are not good enough yet to see the difference. One has to wait for the next-generation EHT (ngEHT) \cite{Ayzenberg:2023hfw}.
\end{enumerate}

Overall, our analysis establishes RHCBH as a theoretically well-motivated and observationally viable alternative to the Kerr black hole. Their difference in its shadow size from the Kerr prediction is controlled by $b$, and its parameters are constrained but not eliminated by the available data.  The bottom line is that RHCBH can still explain the shadows of both M87* and Sgr A*. A large chunk of the parameter space is still allowed, and non-zero quantum corrections are not ruled out. In any case, observations of M87* and Sgr A* are expected to improve our understanding of quantum gravity.

There are many interesting avenues that are amenable for future work. One important extension is to include more realistic accretion flow models to produce complete black hole images that can be compared directly with observations. It is also important to study charged RHCBHs and examine their observational signatures using X-ray observations of hot spots and quasi-periodic oscillations. Another promising direction is to investigate the gravitational-wave ringdown signals and possible echo features of RHCBHs and compare them with data from the LIGO Scientific Collaboration–Virgo Collaboration–KAGRA Collaboration detectors. 
Better telescopes  (ngEHT or next-generation VLBI arrays with longer baselines and higher frequencies) and more precise measurements of black hole masses and distances.

\section*{Data Availability}

We have not generated any original data in due course of this study, nor have any third-party data been analyzed in this article.



\bibliography{RHCBH}
\end{document}